\documentclass[showpacs,preprintnumbers,amsmath,amssymb,floatfix]{revtex4}

\usepackage{color}   
\usepackage{graphicx}
\usepackage{dcolumn}
\usepackage{bm}

\begin{document}

\def\bbox#1{\hbox{\boldmath${#1}$}}
\def\blambda{{\hbox{\boldmath $\lambda$}}}
\def\eeta{{\hbox{\boldmath $\eta$}}}
\def\bxi{{\hbox{\boldmath $\xi$}}}
\def\bzeta{{\hbox{\boldmath $\zeta$}}}
\def\sD{D \!\!\!\!/}

\title{ The Wigner Function of Produced Particles in String Fragmentation}

\author{Cheuk-Yin Wong}

\affiliation{Physics Division, Oak Ridge National Laboratory, 
Oak Ridge, TN\footnote{wongc@ornl.gov} 37831}

\date{\today}

\begin{abstract}

We show that QCD4 with transverse confinement can be approximately
compactified into QCD2 with a transverse quark mass $m_{{}_T}$ that is
obtained by solving a set of coupled transverse eigenvalue equations.
In the limits of a strong coupling and a large number of flavors, QCD2
further admits Schwinger QED2-type bosonized solutions.  We therefore
examine phenomenologically the space-time dynamics of produced
particles in string fragmentation by studying the Wigner function of
produced bosons in Schwinger QED2, which mimics many features of
string fragmentation in quantum chromodynamics. We find that particles
with momenta in different regions of the rapidity plateau are produced
at the initial moment of string fragmentation as a quark pulls away
from an antiquark at high energies, in contrast to classical pictures
of string fragmentation with longitudinal space-momentum-time
ordering.

\end{abstract}


\maketitle


\section{Introduction}

Soft particle production by string fragmentation is an important
process in high energy collisions.  In the strongly-coupled regime in
nucleus-nucleus collisions at RHIC energies, the dynamics of string
fragmentation falls within the realm of non-perturbative quantum
chromodynamics (QCD), and the detailed mechanism of the production
process cannot yet be described from the first principles of QCD.
Phenomenological particle production models based on preconfinement
\cite{Wol80}, parton-hadron duality \cite{Van88} cluster fragmentation
\cite{Odo80}, string-fragmentation \cite{And83}, dual-partons
\cite{Cap78}, the Venus model \cite{Wer88}, the RQMD model
\cite{Sor89}, multiple collision model \cite{Won89}, parton cascade
model \cite{Wan94,Gei92}, color-glass condensate model \cite{McL94},
the AMPT model \cite{Li96}, the Lexus model \cite{Jeo97}, and many
other models have been put forth and are successful in describing some
aspects of the data. These models incorporate some features of QCD,
but their fundamental foundation in non-perturbative QCD is still
lacking. It is of interest to study the physics of particle production
in string fragmentation with QCD-inspired, low-dimensional
non-perturbative models.

Two-dimensional quantum electrodynamics
\cite{Sch62,Low71,Col75,Col76,Cas74,Bjo83,Ili84,Ell87,Fuj89,Won91,Won94}
in one space and one time coordinates (QED2) furnishes an interesting
arena for such a study.  This is a quantum mechanical system in which
a fermion interacts with an antifermion through a linear gauge field
potential in the Coulomb gauge.  It is a system where the Higgs
phenomenon occurs and the global chiral symmetry is spontaneously
broken \cite{Sch62,Low71,Col75,Col76}.  When a positive fermion and a
negative antifermion are separated in such a system, the vacuum is so
polarized that the positive and the negative charges are completely
screened, in a manner similar to the confinement of quarks.  The
quantum field theory of massless fermions interacting with a gauge
field in QED2 can be solved exactly
\cite{Sch62,Low71,Col75,Col76,Cas74,Bjo83}.  The interacting field
turns out to be equivalent to the quantum field theory of free bosons
with a mass.  It was demonstrated by Casher, Kogut and Susskind
\cite{Cas74} that when a fermion separates from an antifermion in the
limit of infinite energies in QED2, the rapidity distribution of the
produced particles will exhibit the property of boost invariance.  For
a finite energy system, the boost-invariant solution turns naturally
into a rapidity plateau, whose width increases with energy as
$\ln(\sqrt{s})$ \cite{Won91,Won94}.

Rapidity plateau distributions of produced particles have indeed been
observed in high-energy $e^+$-$e^-$ annihilation which produces
initially a separating quark and an antiquark
\cite{Aih88,Hof88,Aihnote,Pet88,Abe99,Abr99}.  Rapidity plateau
distributions of produced particles have also been observed in $pp$
collisions at $\sqrt{s}=200$ GeV by the BRAHMS Collaboration
\cite{Yan08}.  The close similarities of the features of empirical
fragmentation data and theoretical QED2 fragmentation results motivate
us to examine in this paper the circumstances in which the
longitudinal dynamics of a transversely-confined QCD4 system can be
approximately compactified into QCD2. We also need to examine further
how multi-flavor QCD2 can be approximated by QED2.

Our renewed interest in the non-perturbative production mechanism
arises from the puzzling observations of the momentum distribution of
particles associated with a near-side jet. In high-energy heavy-ion
collisions, a near-side jet is characterized by the presence of
associated particles within a narrow cone along the trigger particle
direction.  Its characteristics resemble those of a jet in $pp$ and
peripheral heavy-ion collisions.  In addition to this `jet component',
there is an additional `ridge component' of associated particles at
$\Delta\phi$$ \sim$0 with a ridge structure in $\Delta \eta$, where
$\Delta \phi$ and $\Delta \eta$ are the azimuthal angle and
pseudorapidity differences measured relative to the trigger particle
\cite{Ada05,Ada06,Put07,Bie07,Wan07,Bie07a,Abe07,Mol07,Lon07,Nat08,Fen08,
Net08,Bar08,Dau08,Ada08,Mcc08,Che08,Wen08,Tan08,Jia08qm,Lee08}.

In the phenomenon of the ridge associated with the near-side jet, it
is observed that (i) the ridge particle yield increases with the
number of participants, (ii) the ridge yield appears to be nearly
independent of the trigger jet properties, (iii) the baryon to meson
ratios of the ridge particles are more similar to those of the bulk
matter than those of the jet, and (iv) the slope parameter of the
transverse distribution of ridge particles is intermediate between
those of the jet and the bulk matter \cite{Ada05}-\cite{Lee08}. These
features suggest that the ridge particles are medium partons, at an
earlier stage of the medium evolution during the passage of the jet.
The azimuthal correlation of the ridge particle with the jet and the
presence of strong screening suggest that the associated ridge
particle and the trigger jet are related by a collision.  A momentum
kick model was put forth to explain the ridge phenomenon
\cite{Won07,Won07a,Won08,Won08a,Won09}.  The model assumes that a near-side
jet occurs near the surface, collides with medium partons, loses
energy along its way, and fragments into the trigger and its
associated fragmentation products (the ``jet component'').  Those
medium partons that collide with the jet acquire a longitudinal
momentum kick along the jet direction.  They subsequently materialize
by parton-hadron duality as ridge particles in the ``ridge
component''.  They carry direct information on the momentum
distribution of the medium partons at the moment of jet-(medium
parton) collisions.  Based on such a description, the experimental
ridge data of the STAR, PHOBOS, and PHENIX Collaborations
\cite{Ada05,Put07,Wan07,Wen08,Ada08} reveal an early parton momentum
distribution that has a thermal-like transverse momentum distribution
and a rapidity plateau extending to rapidities as large as $|y|\sim
4$.

The early stage of a nucleus-nucleus collision consists of
simultaneous particle production processes involving quarks pulling
away from antiquarks (or $qq$ diquarks) at high energies.  As
experimental data of elementary production processes have a rapidity
plateau structure \cite{Aih88,Hof88,Aihnote,Pet88,Abe99,Abr99,Yan08},
partons at the early stage of the nucleus-nucleus collision can
possess the characteristics of a rapidity plateau.  A rapidity plateau
would also be expected from the theoretical model of string
fragmentation in QED2 \cite{Cas74,Bjo83,Won91,Won94}, the dual-parton
model \cite{Cap78}, the Bjorken hydrodynamics \cite{Bjo83}, the Lund
model \cite{And83}, the Venus model \cite{Wer88}, color-glass
condensate model \cite{McL94}, the Lexus model \cite{Jeo97}, and many
other models.  Thus, the possibility of a rapidity plateau for the
ridge particles should not come as a surprise.  However, the fact that
a parton of large absolute rapidity can occur together with a jet of
central rapidities, as revealed by the PHOBOS data and the momentum
kick model analysis, poses an interesting conceptual question.  For
those medium partons with large magnitude of the longitudinal momentum
to collide with the jet so as to become an associated ridge particle
in the PHOBOS experiment, the partons must be present in the
longitudinal neighborhood of the jet at the moments of the jet-medium
collision, at an early stage of the nucleus-nucleus collision.

In many classical descriptions of particle production processes, a
particle with a large absolute rapidity is associated with a large
separation from the longitudinal origin.  For example, in the Lund
model of string fragmentation \cite{And83}, there is a
momentum-space-time ordering of produced particles in the
fragmentation of a string.  Consider the case in the center-of-mass
system.  Those particles with the smallest absolute rapidity will be
produced nearest to the origin at the earliest times, while those with
the largest absolute rapidity will be produced farthermost from the
origin at the latest times \cite{And83,Won94}.

In Bjorken hydrodynamics \cite{Bjo83}, the spatial configurations are
invariant with respect to a boost in rapidity.  There is a
longitudinal momentum-space ordering of the produced particles.
Particles at the local longitudinal location $z=0$ have rapidity zero
and it is necessary to go to a relatively large absolute longitudinal
coordinate $|z|$ to find another particle with a large absolute
rapidity $|y|$.

In another classical model of Landau hydrodynamics
\cite{Lan53,Won08Lan1,Won08Lan2}, one starts with a fluid initially at
rest.  As the fluid evolves, there is also an ordering of the rapidity
with the longitudinal coordinate and time.  It is necessary to go to
larger absolute longitudinal coordinates at later times to find
particles with larger absolute rapidities.

In these classical pictures of produced particles, there is a
space-momentum-time ordering.  Those partons with large values of
absolute rapidities are not produced at the early stage of the jet
production and therefore cannot be associated with a jet by a
collision.  On the other hand, the PHOBOS data \cite{Wen08} and the
momentum kick model analysis \cite{Won07,Won07a,Won08,Won08a,Won09}
indicate that a central rapidity jet and another parton at a large
absolute rapidity can be associated by a collision.  How does one
resolve the difference of these two seemingly contradicting results?

It should be pointed out that the momentum-space-time ordering of
produced particles in the above classical descriptions is obtained in
final-state-interaction models.  There is however another class of
well-justified initial-state-interaction models, in which the
constituent particles or produced particles are present at or before
the collision process at $t=0$.  Notable examples are the parton model
\cite{Fey69} and the Drell-Yan process \cite{Dre71}, which assume the
presence of constituent partons inside the hadron or in the
quark-antiquark sea before collisions.  The multi-peripheral model
\cite{Che68} assumes the exchange of a tower of reggeons and the
plateau of particles are present before the collision at $t$=0.  The
dipole approach in photon-induced reactions \cite{Huf00} assumes the
fluctuation of the incident photon into a dipole quark-antiquark pair
before its collision with the hadron.

With regard specifically to the dynamics of produced particles in the
fragmentation of a string, it is important to point out that there are
important quantum effects \cite{Jor35,Sch62,Ili84} arising from the
vacuum structure in strongly-coupled non-perturbative quantum field
theories that are beyond the realm of classical considerations.  The
proper space-time dynamics should be based on a quantum field theory
where the quantum effects of the vacuum structure play appropriate
roles.  As is well known, in theories where the fermions are weakly
interacting and can be isolated, the bare vacuum suffices for the
description of the dynamics in a weak-coupling perturbation theory.
In contrast, however, in our systems with strong coupling between
fermions, such as in massless QED2 and similar field theories, the
vacuum of the filled Dirac sea is the appropriate vacuum which
participates in the dynamics.  The self-consistent dynamics of the
filled Dirac sea in the non-perturbative response to the presence of
gauge field and fermion disturbances lead to peculiar quantum effects
of the axial anomaly \cite{Jor35,Col75,Col76,Ili84}, the confinement
of the fermion \cite{Sch62}, and the appearance of massive bosons
\cite{Sch62}.  The particle production process in string fragmentation
should therefore be treated in a non-perturbative quantum field theory
framework.

While a complete non-perturbative theory based on full quantum
chromodynamics is desirable, it is however not available.  Simplifying
approximations are needed.  We shall show that a system of
transversely-confined QCD4 can be approximately compactified to a
system of QCD2 with a transverse quark mass $m_{{}_T}$ that is
obtained from a set of coupled transverse eigenvalue equations.  We
shall show that in the limit of strong coupling and large number of
flavors, a multi-flavor massless QCD2 admits Schwinger QED2-type
solutions with an effective coupling constant that depends on the
number of flavors.  In the absence of rigorous non-perturbative QCD4
solutions, we are therefore justified to examine phenomenologically
the dynamics of the system in the solvable quantum field theory of
QED2, on accounts of its close theoretical connection with the
transversely-confined QCD4 and its desirable properties of the proper
high-energy rapidity plateau behavior, confinement, charge screening,
and chiral symmetry breaking.

The space-time dynamics of produced particles in string fragmentation
can be obtained by evaluating the Wigner function of the produced
particles when a strongly-coupled fermion separates from an
antifermion at high energies in QED2.  In contrast to the classical
description of particle production, we shall find that produced
particles with different momenta in different regions of the rapidity
plateau are present at the moment when the overlapping
fermion-antifermion pair begin to separate.

The space-time dynamics of partons in the central rapidity region has
been obtained in the color-glass condensate model \cite{McL94}.  In
contrast to the strongly-coupled non-perturbative field theory
discussed here, the color-glass condensate model considers the region
of $x\ll 1$ and $p_T \gg \Lambda_{\rm QCD}$ in the weak-coupling limit
of isolated quarks and gluons.  The space-time dynamics has been
evaluated in the boost-invariant limit of infinite energies with the
valence quark in the no-recoil approximation.  The description is
appropriate for the production of particles in the central rapidity
region with large transverse momenta.  In the present case we wish to
examine the space-time dynamics of the production of particles with
large (pseudo)rapidities ($|\Delta \eta| \sim 4$) and small transverse
momenta ($p_t > 0.035$ GeV), such as those ridge particles detected by
the PHOBOS Collaboration \cite{Wen08}, in a system of finite energy.
These particle are in the region of large $x$ and small $p_t \ll
\Lambda_{\rm QCD}$, and the dynamics is within the realm of strong
coupling.  It is therefore appropriate to investigate here the
space-time dynamics of produced particles in string fragmentation in a
strongly-coupled non-perturbative quantum field theory.

This paper is organized as follows.  In Section II, we spell out
explicitly the circumstances in which a transversely-confined QCD4 can
be approximately compactified to a massive QCD2 of fermions with a
transverse mass $m_T$ which is the eigenvalue of transverse
confinement.  In Section III, we study how the multi-flavor massless
QCD2 and QED2 are related in the limit of a strong coupling and a
large number of flavors $N_f$.  The dynamics can then be considered as
those of QED2 with an effective coupling constant that is proportional
to $\sqrt{N_f}$.  In Section IV, we review the particle production
processes in massless QED2 and show how the Wigner function can be
obtained from the initial fermion current. In Section V, we give an
explicit example in the fragmentation of a string with a finite energy
and compare the experimental rapidity distribution data in high energy
$e^+$-$e^-$ annihilation with the theoretical rapidity distribution.
In Section VI, The Wigner function of the produced particles is
evaluated as a function of time to study the space-time dynamics of
produced particles in the fragmentation of a string in high energy
$e^+$-$e^-$ annihilation.  In Section VII, we generalize the results
for a single string to the case of many identical strings at different
locations and show how the Wigner function can be evaluated.  In
Section VIII, we give numerical examples of the Wigner function in the
fragmentation of identical strings.  The Wigner function exhibits the
effects of interference between different strings.  In Section IX, we
discuss schematically the space-time dynamics in the case of the
fragmentation from many independent string in the context of
high-energy heavy-ion collisions.  We present our discussions and
conclusions in Section X.

\section{Approximate Compactification of Transversely-Confined QCD4 to QCD2}

Under certain appropriate conditions, quantum chromodynamics in 4
dimensions (QCD4) can be approximately compactified to quantum
chromodynamics in 2 dimensions (QCD2).  It is useful to describe the
circumstances and assumptions leading to this simplification.
Previously, t'Hooft showed that in the limit of large $N_c$ with fixed
$g^2N_c$ in single-flavor QCD4, planar diagrams with quarks at the
edges dominate, whereas diagrams with the topology of a fermion loop
or a wormhole are associated with suppressing factors of $1/N_c$ and
$1/N_c^2$, respectively \cite{tho74a}.  In this case a simple-minded
perturbation expansion with respect to the coupling constant $g$
cannot describe the spectrum, while the $1/N_c$ expansion may be a
reasonable expansion, in spite of the fact that $N_c$ is equal to 3
and is not very big.  The dominance of the planar diagram allows one
to consider QCD in one space and one time dimensions (QCD2) and the
physics resemble those of the dual string or a flux tube, with the
physical spectrum of a straight Regge trajectory \cite{tho74b}.  Since
the pioneering work of t'Hooft, the properties of QCD2 systems have
been investigated by many workers
\cite{tho74a,tho74b,Fri93,Dal93,Fri94,Arm95,Kut95,Gro96,Abd96,Dal98,Arm99,Eng01,Tri02,Abr04,Li87,Wit84}.
The flux tube picture of the longitudinal dynamics is further
phenomenologically supported by hadron spectroscopy \cite{Isg85} and
the limiting average transverse momentum of produced particles in
high-energy hadron collisions \cite{Gat92}.

The description of the QCD4 dynamics as a flux tube and the dominance
of the longitudinal motion over the the transverse motion in string
fragmentation provide the motivation to compactify, at the least
approximately, the transversely-confined QCD4 to QCD2 and QED2.  Such
a link was presented briefly in \cite{Won08a}.  We would like to
reiterate the derivation further to show how the coupling constant of
a transversely-confined QCD4 can be related to the coupling constant
in QCD2 in such a compactification.

We start with QCD4 in which the dynamical variables are the quarks
fields $\psi^{ab}$ and the gauge fields $A_\nu^a $, where $a$ and $b$
are the color and flavor indices respectively, and $\nu=0,1,2,3$. The
color and flavor indices will not be displayed explicitly except when
they are needed.  The gauge fields are coupled to the quark fields
which are in turn coupled to the gauge fields, resulting in a coupled
system of great complexities.  The transverse confinement of the flux
tube can be represented by quarks moving in a nonperturbative scalar
field $m({\bf r})$, part of which may contain the flavor-dependent
quark rest masses.  The equation of motion of a quark field $\psi$ is
\begin{eqnarray}
\label{quark1}
\left \{ i \sD - m({\bf r}) \right \} \psi =0
\end{eqnarray}
where 
\begin{eqnarray}
i\sD = \gamma^\mu \Pi_\mu ,
\end{eqnarray}
\begin{eqnarray}
\Pi_\mu = p_\mu + g A_\mu,
\end{eqnarray}
\begin{eqnarray}
A_\mu=\tau^a A_\mu^a,
\end{eqnarray}
and $\tau^a$ are the generators of the color SU($N_c$) group.
The equation of motion for the gauge field $A_\mu$ is
\begin{eqnarray}
\label{Max4}
D_\mu F^{\mu \nu} = g {\bar \psi} \gamma^\nu {\bf \tau} \psi,
~~~~~~ \mu,\nu=0,1,2,3,
\end{eqnarray}
where
\begin{eqnarray}
\label{F1}
F_{\mu \nu} = \partial_\mu A_\nu - \partial_\nu A_\mu -i g [A_\mu, A_\nu],
\end{eqnarray}
\begin{eqnarray}
\label{F2}
D_\mu F^{\mu \nu} = \partial_\mu F^{\mu \nu} -i g [A_\mu, F^{\mu\nu}],
\end{eqnarray}
\begin{eqnarray}
\label{F3}
F_{\mu \nu}=\tau^a F_{\mu \nu}^a.
\end{eqnarray}
The fermion current that is the source of the gauge field is
\begin{eqnarray}
\label{F4}
j^ \nu = g {\bar \psi} \gamma^\nu {\bf \tau}  \psi.
\end{eqnarray}

We shall consider the problem of string fragmentation in which the
momentum scale for the longitudinal dynamical motion of the source
current is much greater than the momentum scale for the transverse
motion, $|v^3|\gg |v^1|,|v^2|$, where ${\bf v}$ is a typical source
velocity.  In the Lorentz gauge, the associated gauge field $A^\mu$ is
proportional to $(1,{\bf v})$.  Under the dominance of the
longitudinal motion over the transverse motion in string
fragmentation, $A^1$ and $A^2$ can be neglected in comparison with the
magnitudes of $A^0$ or $A^3$.  For the flux tube configuration, it is
further reasonable to assume that the gauge fields $A^0$ and $A^3$
inside the tube depend only weakly on the transverse coordinates ${\bf
r}=(x^1,x^2)$.  It is meaningful to investigate these fields inside
the tube by averaging them over the transverse profile of the flux
tube.  After such an averaging, $A^0$ and $A^3$ can be considered as a
function of $(x^0,x^3)$ only.  As a consequence, the equation of
motion for the quarks become
\begin{eqnarray}
\label{quarkp}
\left \{ \gamma^0 \Pi_0 + \gamma^1 p_1 + \gamma^2 p_2 +
\gamma^3 \Pi_3 - m({\bf r}) \right \} \psi = 0,
\end{eqnarray}
where $\Pi_0$ and $\Pi_3$ have been approximated to be independent of
the transverse coordinates.  To separate out the longitudinal and
transverse degrees of freedom, we expand the quark field in terms of
spinors $\mu_i$ \cite{Won91}
\begin{eqnarray}
\label{quark}
\psi(x)=(G_1 \mu_1 - G_2 \mu_2) f_+  + ( G_1 \mu_3  +G_2 \mu_4) f_-  
\end{eqnarray}
where $G_{1,2}$ are functions of the transverse coordinate ${\bf r}$,
 $f_{\pm}$ are functions of $( x^0, x^3)$, and the spinors have been
 chosen to be eigenstates of the gamma matrix $\alpha$ with $\alpha
 \mu_{1,2}=\mu_{1,2}$ and $\alpha \mu_{3,4}=-\mu_{3,4}$ \cite{Won91},
\begin{eqnarray}
\mu_1=\left ( \begin{matrix} 1\\
                     0\\
                     1\\
                     0
               \end{matrix}
      \right ),
~~~~
\mu_2=\left ( \begin{matrix} 0\\
                     1\\
                     0\\
                     -1
               \end{matrix}
      \right ),
~~~~
\mu_3=\left ( \begin{matrix} 1\\
                     0\\
                    -1\\
                     0
               \end{matrix}
      \right ),
{\rm ~~~and}~~~~
\mu_4=\left ( \begin{matrix} 0\\
                     1\\
                     0\\
                     1
               \end{matrix}
      \right ).
\end{eqnarray}
Working out the Dirac matrices, we obtain the set of coupled equations
of motion for the quarks
\begin{subequations}
\begin{eqnarray}
\label{a}
-m({\bf r}) G_1 f_+ +(-p^1+ip^2) G_2 f_+ +(\Pi^0+\Pi^3)G_1f_-&=&0,\\
\label{b}
-(p^1+ip^2) G_1 f_+ + m({\bf r}) G_2 f_+ +(\Pi^0+\Pi^3)G_2f_-&=&0,\\  
\label{c}
(\Pi^0-\Pi^3)G_1f_+- m({\bf r}) G_1 f_-  -(p^1-ip^2) G_2 f_- &=&0,\\
\label{d}
-(\Pi^0-\Pi^3)G_2f_++(p^1+ip^2) G_1 f_- - m({\bf r}) G_2 f_- &=&0 .
\end{eqnarray}
\end{subequations}
Applying $\Pi^0-\Pi^3$ on (\ref{a}) and using (\ref{c}) and (\ref{d}), we get
\begin{eqnarray}
[(\Pi^0-\Pi^3)(\Pi^0+\Pi^3)- m^2({\bf r})
-{\bf p}_T^2]G_1f_- = [(-p^1+ip^2)m({\bf r})]G_2f_- .
\end{eqnarray}
By the method of the separation of variables, we can introduce the
eigenvalue $m_T^2$ such that the above equation can be separated into
an equation for the transverse coordinates and an equation for the
longitudinal and time coordinates,
\begin{eqnarray}
\label{f}
[{\bf p}_T^2
+ m^2({\bf r})-m_T^2]G_1({\bf r}) = [(p^1-ip^2)m({\bf r})]G_2({\bf r}),
\end{eqnarray}
\begin{eqnarray}
\label{h}
[(\Pi^0-\Pi^3)(\Pi^0+\Pi^3)-m_T^2] f_-(x^0,x^3) =0.
\end{eqnarray}
By applying $\Pi^0-\Pi^3$ on (\ref{b}) and using (\ref{c}) and
(\ref{d}), we can separate out the same longitudinal equation but the
following transverse equation,
\begin{eqnarray}
\label{g}
[{\bf p}_T^2
+ m^2({\bf r})-m_T^2]G_2({\bf r}) = -[(p^1+ip^2)m({\bf r})]G_1({\bf r}).
\end{eqnarray}
Similarly, by applying $\Pi^0+\Pi^3$ on (\ref{c}) and using (\ref{a})
and (\ref{b}), we can separate again the same transverse equation
(\ref{f}) for $G_1$, and the following longitudinal equation
\begin{eqnarray}
\label{i}
[(\Pi^0+\Pi^3)(\Pi^0-\Pi^3)-m_T^2] f_+(x^0,x^3) =0.
\end{eqnarray}

The set of four coupled equations of (\ref{a})-(\ref{d}) from Eq.\
(\ref{quark1}) (or (\ref{quarkp})) for the quark therefore becomes the
set of four equations of (\ref{f})-(\ref{i}).  Eqs.\ (\ref{f}) and
(\ref{g}) provide the equations for the solution of the eigenvalue
$m_T^2$ under the boundary condition that the functions $G_1$ and
$G_2$ are confined, with vanishing probabilities at $|{\bf r}| \to
\infty$.  Note that Eqs.\ (\ref{f}) and (\ref{g}) is applicable not
only to an azimuthally symmetric transverse potential but also to an
azimuthally-asymmetric potential $m({\bf r})$.  For an azimuthally
symmetric transverse potential $m(r)$, we can further write the
transverse wave functions as
\begin{subequations}
\begin{eqnarray}
G_1(r,\phi)&=& g_1(r)e^{i \nu \phi}, \\
G_2(r,\phi)&=& g_2(r)e^{i (\nu+1) \phi}. 
\end{eqnarray}
\end{subequations}
Eqs.\ (\ref{f}) and (\ref{g}) then become a coupled set of equations for
$g_1(r)$ and $g_2(r)$ \cite{Not2},
\begin{subequations}
\begin{eqnarray}
\label{dpdr1}
\left [ {\bf p}_{{}_T}^2(\nu) +m^2(r) - m_{{}_T}^2 \right ] g_1(r) &=&-i
\frac{\partial m(r)}{\partial r} g_2(r), \\
\label{dpdr2}
\left [ {\bf p}_{{}_T}^2(\nu) +m^2(r) - m_{{}_T}^2 \right ] g_2(r) &=&i
\frac{\partial m(r)}{\partial r}  g_1(r), 
\end{eqnarray}
\end{subequations}
where
\begin{eqnarray}
{\bf p}_{{}_T}^2(\nu) = -\frac{1}{r}\frac{\partial}{\partial r} 
r\frac{\partial}{\partial r} +\frac{\nu^2}{r^2}.  
\end{eqnarray}

In solving the transverse eigenvalue equations (\ref{f}) and
(\ref{g}), (or (\ref{dpdr1}) and (\ref{dpdr2})) with the boundary
condition of transverse confinement, an eigenstate is characterized by
transverse quantum numbers and the corresponding eigenvalue $m_T$
depends on these quantum numbers.  Thus, the transverse quark mass
$m_{{}_T}$ will depend on the transverse quantum state of the quark in
the flux tube.  Excitation of the quark in the flux tube will lead to
a greater transverse mass.  There is thus an additional degree of
freedom for the transverse quark mass.  The excitation of the
transverse state will lead to a broader width of the transverse
momentum distribution for the quark.  Composite objects formed by
pairing a quark with an antiquark will likewise acquire a broader
transverse momentum distribution \cite{Gat92}.

Eqs. (\ref{h}) and (\ref{i}) are the longitudinal equations for a
quark in the two-dimensional space-time $(x^0,x^3)$.  They can be
rewritten as a Dirac equation in the two-dimensional space-time as
follows.  We introduce a two-dimensional quark spinor
\begin{eqnarray}
\label{quarkf}
\psi_{\rm 2D} = 
\left ( \begin{matrix} f_+\\
                       f_-
        \end{matrix}
      \right )
\end{eqnarray}
and use two-dimensional gamma matrices \cite{Wit84,Abd96}
\begin{eqnarray}
\gamma_{\rm 2D}^0 = 
\left ( \begin{matrix} 0 & 1\\
                       1 & 0
        \end{matrix}
      \right ),
~~~~
\gamma_{\rm 2D}^3 = 
\left ( \begin{matrix} 0 & -1\\
                       1 & 0
        \end{matrix}
      \right ),
~~~~
\gamma_{\rm 2D}^0 \gamma_{\rm 2D}^3 = \gamma_{\rm 2D}^5=
\left ( \begin{matrix} 1 & 0\\
                       0 & -1
        \end{matrix}
      \right ).
\end{eqnarray}
Then Eqs. (\ref{h}) and (\ref{i}) can be written as
\begin{eqnarray}
\label{predir}
\left \{ (\Pi^0)^2 - (\Pi^3)^2 + \gamma_{\rm 2D}^3 \gamma_{\rm 2D}^0 
[\Pi^3, \Pi^0] - m_T^2 
\right \} \psi_{\rm 2D} =0.
\end{eqnarray}
This same equation (\ref{predir})  can also be obtained from
\begin{eqnarray}
\label{dir4}
\left \{ \gamma_{\rm 2D}^0\Pi_0+ \gamma_{\rm 2D}^3\Pi_3- m_T \right \} \psi_{\rm 2D}
= \left \{ \gamma_{\rm 2D}^0(p_0 + g A_0)+ \gamma_{\rm  2D}^3(p_3+gA_3)- m_T \right \} \psi_{\rm 2D} =0,
\end{eqnarray}
which is the equation of motion for a quark in two-dimensional gauge fields
of $A_0$ and $A_3$.

Note that the above equation depends on $g_{\rm 4D} A_{\mu({\rm 4D})}$
where we have added the subscript `4D' to $g$ and $A_\mu$ to indicate
that the coupling constant is the coupling constant in QCD4, and the
gauge fields are from the Maxwell equation (\ref{Max4}) in the
four-dimensional space-time.  In the two-dimensional space-time of QCD2,
the corresponding equations of motion for the quark and the gauge
fields are
\begin{eqnarray}
\label{dir2}
\left \{ \gamma_{\rm 2D}^0(p_0 + g_{\rm 2D} A_{0({\rm 2D})})+ 
\gamma_{\rm  2D}^3(p_3+g_{\rm 2D} A_{3({\rm 2D})})
- m_T \right \} \psi_{\rm 2D} =0,
\end{eqnarray}
and 
\begin{eqnarray}
\label{Max2}
D_\mu F_{\rm 2D}^{\mu \nu} 
= g_{\rm 2D} {\bar \psi}_{\rm 2D} \gamma_{\rm 2D}^\nu{\bf \tau}  \psi_{\rm
2D}   =j_{\rm 2D} ^\nu,~~~~~~{\mu \nu=0,3},
\end{eqnarray}
where $F_{\rm 2D}^{\mu \nu}$ are given by Eqs.\ (\ref{F1})-(\ref{F3})
with $A_{\nu({\rm 4D})}$ replaced by $A_{\nu({\rm 2D})}$ and $g_{\rm
4D}$ by $g_{\rm 2D}$.

To cast Eq.\ (\ref{dir4}) in the form of (\ref{dir2}) we can identify
\begin{eqnarray}
g_{\rm 4D}A_{\mu ({\rm 4D})}= g_{\rm 2D}A_{\mu ({\rm 2D})}.
\end{eqnarray}  
To cast Eq.\ (\ref{Max4}) in the form of Eq.\ (\ref{Max2}), we need a
relationship between $ j_{\rm 2D} ^\nu$ and $j_{\rm 4D} ^\nu$.  With
the quark field as given by (\ref{quark}) and (\ref{quarkf}), we have
\begin{eqnarray}
j_{\rm 4D}^\nu &=& g_{\rm 4D} {\bar \psi}_{\rm 4D} \gamma_{\rm 4D}^\nu
{\bf \tau} \psi_{\rm 4D} \nonumber\\ 
&=& g_{\rm 4D} (|G_1({\bf r})|^2 + 
|G_2({\bf r})|^2) {\bar \psi}_{\rm 2D} \gamma_{\rm 2D}^\nu {\bf
  \tau} \psi_{\rm 2D}, \nonumber\\ 
&=& (g_{\rm 4D}/g_{\rm 2D}) (|G_1({\bf r})|^2 + |G_2({\bf r})|^2) j_{\rm 2D}^\nu.
\end{eqnarray}
This indicates that the $j_{\rm 4D}^\nu$ current depends also on the
transverse spatial density distribution.  As our focus is in the
longitudinal dynamics for high-energy string fragmentation, it
suffices to average the quark transverse density $|G_1({\bf r})|^2 +
|G_2({\bf r})|^2$ over the flux tube transverse profile to relate
$j_{\rm 4D}$ approximately with $j_{\rm 2D}$ as
\begin{eqnarray}
\langle j_{\rm 4D} ^\nu \rangle_{{}_T} =(g_{\rm 4D}/g_{\rm 2D}) \langle (|G_1({\bf r})|^2
+ |G_2({\bf r})|^2) \rangle_{{}_T} j_{\rm 2D}^\nu,
\end{eqnarray}
 where the transverse averaging of ${\cal O}$ is defined as
\begin{eqnarray}
\langle {\cal O} \rangle_{{}_T}= \int d{\bf r} {\cal O}
 (|G_1({\bf r})|^2 + |G_2({\bf r})|^2).
\end{eqnarray}
The source term of the Maxwell equation involves a coupling constant
and the gauge field quantity in the Dirac equation also involves the
coupling constant.  The equations of motion of the quarks and gluons
(\ref{dir2}) and (\ref{Max4}) can be cast in the forms of (\ref{dir2})
and (\ref{Max2}) by transversely averaging Eq. (\ref{Max4}) over the
profile of the flux tube and by relating the coupling constants by the
following renormalization
\begin{eqnarray}
g_{\rm 2D}^2 = g_{\rm 4D}^2 \langle (|G_1({\bf r})|^2 + |G_2({\bf
r})|^2)\rangle_{{}_T}.
\end{eqnarray}
We can get an estimate of the relation between $g_{\rm 2D}$ and
$g_{\rm 4D}$ by considering the case of a uniform transverse flux tube
profile with a transverse radius $R_T$,
\begin{eqnarray}
(|G_1({\bf r})|^2 + |G_2({\bf
r})|^2) \sim \Theta(R_T - |{\bf r}|)/\pi R_T^2.
\end{eqnarray}
We have then 
\begin{eqnarray}
\label{est}
g_{\rm 2D}^2 = g_{\rm 4D}^2/(\pi R_T^2)=4 \alpha_s /R_T^2.
\end{eqnarray}
If we consider the case of $e^+$-$e^-$ annihilation at $\sqrt{s}=29$
GeV, the average $p_T^2$ of produced pions is 0.255 GeV$^2$
\cite{Pet88} corresponding to a flux tube radius of order
\begin{eqnarray}
R_T \sim \hbar/\sqrt{\langle p_T^2\rangle} \sim 0.4 {\rm ~~fm}.
\end{eqnarray}
A  typical case of $\alpha=0.4$ and a flux tube radius
of $R_T=0.4$ fm then gives 
\begin{eqnarray}
g_{\rm 2D}^2 = 0.40 {\rm ~GeV}^2.
\end{eqnarray}
On the other hand, our earlier comparison relates $g_{\rm 2D}^2$ with
the string tension coefficient $b$ in the linear potential is
\cite{Won08a}
\begin{eqnarray}
g_{\rm 2D}^2 = 2 b,
\end{eqnarray}
which also give $g_{\rm 2D}^2\sim 0.4$ GeV$^2$, for a string tension
of $b=0.2$ GeV$^2$ = 1 GeV/fm.  The estimate of the QCD2 coupling
constant given above in Eq.\ (\ref{est}) is therefore consistent with
the string tension coefficient.  This relationship also gives a
relation between the string tension $b$ and other physical quantities,
\begin{eqnarray}
b=2 \alpha_s /R_T^2.
\end{eqnarray}
Thus, with the many approximations outlined above, a
transversely-confined QCD4 can be compactified as massive QCD2 with a
transverse quark mass $m_T$ obtained by solving the transverse
eigenvalue equations (\ref{f}) and (\ref{g}).

Our explicit formulation to compactify a four-dimensional space-time
in a flux tube to a two-dimensional space-time serves to provide a
great simplifications which allows us to examine the dynamics of
systems dominated by the dynamics along the longitudinal direction, as in
high energy string fragmentation.  The transverse degrees of freedom
is subsumed by the use of the transverse quark mass $m_T$, and the
longitudinal and the transverse degrees of freedom are decoupled.  If
needed subsequently, the transverse properties may be approximately
restored by using the transverse eigenfunctions and their
corresponding transverse momentum distributions of the quarks and
antiquarks.  For example, while we get the rapidity distribution of
the produced bosons, its transverse distribution may be considered by
treating the boson as a composite object made up of a quark and an
antiquark, for each of which the transverse momentum distribution can
be obtained from the transverse eigenfunctions $G_1({\bf r})$ and
$G_2({\bf r})$, as discussed previously in \cite{Gat92}.

\section{ Relation between QCD2 and QED2} 

In the last section we have shown that the transversely-confined QCD4
can be approximately compactified to QCD2 with a transverse
quark mass determined by the transverse confinement of the quarks
within the flux tube.  We can examine the flavor degrees of freedom and
write down the flavor index $b$ in the QCD2 equations of motion
(\ref{dir2}) and (\ref{Max2}).  In the general case when the
transverse mass $m_T^b$ depends on the flavor (as the current quark
masses may be flavor-dependent), the QCD2 equation of motion for the
quark is
\begin{eqnarray}
\label{dirF21}
\left \{ \gamma^0 (p_0 + g_{\rm 2D} A_{0})+ 
\gamma^3(p_3+g_{\rm 2D} A_{3})
- m_T^b \right \} \psi^{b} =0,    {~~~~~ b=1,...,N_f},
\end{eqnarray}
and the Maxwell equation for the gauge field is
\begin{eqnarray}
\label{MaxF21}
D_\mu F^{\mu \nu} =\sum_{b=1}^{N_f} 
g_{\rm 2D} {\bar \psi}^b \gamma^\nu {\bf \tau} \psi^b,   ~~~~\nu,\mu=0,3, 
\end{eqnarray}
where the subscripts `2D' of $\gamma$, $\psi$, $A_\nu$, and $F^{\mu
\nu}$ have been henceforth omitted, for brevity of notation.

We idealize to the case of a system with flavor symmetry so that the
transverse masses and the quark wave functions of the underlying
system are independent of the flavor.  Then the gauge field equation
(\ref{MaxF21}) becomes
\begin{eqnarray}
\label{MaxF23}
D_\mu F^{\mu \nu} =N_f g_{\rm 2D} {\bar \psi}^b \gamma^\nu
{\bf \tau} \psi^b,  ~~~~({\rm no~sum~in~}b).
\end{eqnarray}
As the strength of the source of the gauge field increases with $N_f
g_{\rm 2D}$, the case of a large flavor number $N_f$ and a large
coupling constant is expected to be in the realm of strong coupling.
In this case, the multi-flavor QCD2 can be best investigated by
bosonization as it display a theory of a scalar particle with weak
self-interactions.  The bosonized massive QCD2 gives rise to a
sine-Gordon equation with periodic symmetry that does not yield simple
solutions.  In this case of strong coupling and in high energy
processes for which the energy of the system is much greater than the
transverse masses, it is reasonable to consider massive QCD2 in the
massless limit and treat the mass as a perturbation in
the mass-perturbation theory \cite{Col76,Ada03}.

Accordingly, we can consider the presence of a perturbation of the
gauge field $A^{\nu a}$ ($\nu=0,3$) in massless QCD2 and can choose
the perturbation to be along the direction of a color component $a$.
The corresponding field strength tensor $F^{\mu \nu, a}$ will also be
along the $a$ direction and commutes with $A^{\nu a}$. The equation of
motion for the gauge field (\ref{MaxF23}) becomes the Abelian Maxwell
equation
\begin{eqnarray}
\label{Max5}
\partial_\mu [
 \partial^\mu A^{\nu a} - \partial^{\nu} A^{\mu a} ]
=  N_f j^{\nu,ab}.
\end{eqnarray}
where
\begin{eqnarray}
j^{\nu,ab}
=  g_{\rm 2D} {\bar \psi}^b \gamma^\nu {\tau}^a \psi^b 
~~~~({\rm no~sum~in~}b).
\end{eqnarray}
Upon taking into account gauge invariance and the singularity of the
Greens function, the induced current $j^{\nu, ab}$ for this Abelian
case of massless quarks is related to the gauge field perturbation by
\cite{Sch62,Won94}
\begin{eqnarray}
j^{\nu, ab}= 
-\frac{g_{\rm 2D} ^2}{\pi} \left [ A^{\nu a} - 
\partial^\nu \frac{1}{\partial ^\lambda \partial_\lambda}
\partial_\mu A^{\mu a} \right ].
\end{eqnarray}
The above Abelian Maxwell equation (\ref{Max5}) is satisfied if
\begin{eqnarray}
\label{KL}
-\square A^{\nu a} - \frac{N_f g_{\rm 2D}^2}{\pi} A^{\nu a}=0.
\end{eqnarray}
The equation of motion for the gauge field perturbation then becomes
the Klein-Gordon equation for a boson with a mass square given by
\begin{eqnarray}
\label{mass}
M^2=N_f g_{\rm 2D}^2 /\pi,
\end{eqnarray}
as obtained previously by Frischmann, Sonnenschein, Trittmann and
their collaborators, and confirmed by numerical calculations
\cite{Arm95,Arm99,Tri02,Not1}.  These authors argue further that
considerations such as given above for the classical equations of
motion in massless QCD2 is applicable especially in the large $N_f$
limit, for which the partition function is dominated by the classical
configuration in the functional integral.  Investigation in this limit
captures the quantum behavior of the theory.

For each color index $a$ of $A^{\nu a}$, there is a massive
Klein-Gordon equation (\ref{KL}).  The massive bosons are therefore
($N_c^2-1$)-fold degenerate, which is the degeneracy of the gluon.
The bosons are the quanta of the excitation arising from the
perturbation the color gauge fields $A^{\nu a}$; they have the color
octet property of the underlying gauge field
perturbations. Furthermore, the coupling constant $g_{\rm 2D}$ is
proportional to $1/R_{{}_T}$ as given by Eq.\ (\ref{est}). The boson
mass $M$ as given by Eq.\ (\ref{mass}) is proportional to
$1/R_{{}_T}$.  A boson mass of this magnitude is also expected for a
gluon in the four-dimensional space-time confined to a flux tube of
radius $R_{{}_T}$.  Based on the above, the massive boson states are
effectively gluons because (i) it has the proper degeneracy of the
gluon, (ii) arises from the excitation of the color gauge fields
$A^{\nu a}$, and (iii) it is of the right order of magnitude of the
gluon mass as in QCD4 space-time in a confined flux tube of radius
$R_{{}_T}$.

One can keep track of the degrees of freedom from a different
perspective as one compactifies QCD4 into QCD2 and QED2.  In the
asymptotic freedom regime of QCD4, light quarks and gluons are free
and essentially massless at short distances.  However, for large
distance behavior in the transversely-confined QCD4 in a flux tube,
gluons and quarks are confined in the tube and they acquire a mass of
order $\hbar/R_T$ corresponding to the zero-point energy in the
confining tube.  When transversely-confined QCD4 is compactified into
QCD2, the degree of freedom are the massive quarks and the gauge
fields $A^0$ and $A^3$, where one of the gauge fields can be
eliminated by fixing the gauge and the other can be written in terms
of the fermion currents.  It may appear at first that the gluon
degrees of freedom have disappeared. However, as we note above, the
excitation of the gauge fields $A^{\nu a}$ leads to massive
color-octet boson states with the properties of confined and massive
gluons.  Thus, the gluons in QCD4 reappear as color bosons in the
excitation of the gauge field.  These massive bosons can also be
considered as excitations accompanied by the coupling of
quark-antiquark pairs in the color-octet configuration. The excitation
of the $A^{\nu a}$ gauge fields producing one of these bosons leaves
the remaining fermions in the corresponding color-octet configuration
such that the combined system of the gluons and quarks is
color-neutral.

What is the fate of these color bosons (or gluons) in the picture of
string fragmentation and jet-parton interaction as examined in
\cite{Won07,Won07a,Won08,Won08a,Won09}?  One can envisage the picture
that as they emerge and decay into the detected colorless particles,
the color of these color scalar bosons can be evaporated with the
emission of soft gluons that carry the color away from one boson to
another neighboring boson until all bosons and remaining fermion
composite objects become colorless.  Such a color evaporation picture
then leads to the model of parton-hadron duality \cite{Van88} that
have been extensively used in the discussion of high-energy collision
processes as for example in the Glassma model \cite{McL94}.

The bosonization of multi-flavor QCD2 in the large $N_f$ limit
indicates that the QCD2 can be approximated as QED2 with three
essential modifications \cite{Arm95,Arm99,Tri02,Abr04}.  Firstly, in
the limit of massless QCD2, the boson mass square is given by $g_{\rm
2D}^2 N_f/\pi$ with an additional factor of $N_f$, and this massive
boson state possesses color and has the degeneracy of
$(N_c^2-1)$. Secondly, the non-zero transverse quark mass in massive
multi-flavor QCD2 leads to a mass term which gives rise to
interactions between the color bosons.  In addition to modifying the
particle spectrum \cite{Ada03}, the interactions arising from the
non-zero transverse quark mass will affect the dynamics of particles after
production.  Thirdly, the magnitude of the fermion current generating
the excitation of the gauge field increases linearly with the flavor
number $N_f$.  As a consequence, the number of produced boson
particles should also increase linearly with the flavor number $N_f$.

While the spectrum is known in the limit of large $N_f$ and $N_c$ for
massless QCD2 \cite{Arm95,Arm99,Tri02,Abr04}, the spectrum for the
realistic case of $N=3$ and $N_f\sim 3$ in massive multi-flavor QCD2
has not been obtained.  We shall carry out a phenomenological
fragmentation study by following the Abelian bosonization of QCD2 to
QED2 and shall equate the mass $m$ of the boson in QED2 as the
experimental transverse mass of detected pions in our
analysis.  We shall also likewise treat the QCD fragmentation problem
as a single flavor QED2 problem with the multiplicity of produced
particles modified by the multiplicative factor of $N_f$.

We note in passing that we can carry out an additional coupling
constant renormalization for a multi-flavor QCD2 with flavor symmetry
by identifying
\begin{eqnarray}
\label{eq44}
g_{\rm 2D}' = g_{\rm 2D} \sqrt{N_f}.
\end{eqnarray}
We introduce rescaled gauge field $A_\mu'$ related to $A_\mu$ by
\begin{eqnarray}
g_{\rm 2D}' A_\mu' = g_{\rm 2D} A_\mu.
\end{eqnarray}
Then Eqs.\ (\ref{dirF21}) and (\ref{MaxF23}) becomes
\begin{eqnarray}
\label{dirF2b}
\left \{ \gamma^0 (p_0 + g_{\rm 2D}' A_{0}')+ 
\gamma^3(p_3+g_{\rm 2D}' A_{3}')
- m_T^b \right \} \psi^{b} =0;    {~~~~~ b=1,...,N_f},
\end{eqnarray}
and the Maxwell equation for the gauge field becomes
\begin{eqnarray}
\label{MaxF2b}
D_\mu (F')^{\mu \nu} =
g_{\rm 2D}' {\bar \psi}^b \gamma^\nu {\bf \tau} \psi^b ~~~~({\rm no~sum~in~}b).
\end{eqnarray}
The above set of equations (\ref{dirF2b}) and (\ref{MaxF2b}) is the
same as the set of single-flavor QCD2 equations with an effective coupling
constant $g_{\rm 2D}'$. The dynamics of the flavor-symmetric
multi-flavor QCD2 is therefore the same as a single-flavor QCD2 with a
renormalized coupling constant $g_{\rm 2D}'$ as given by $g_{\rm 2D}
\sqrt{N_f}$ in (\ref{eq44}).

\section{ Wigner function of produced particles}

After discussing the relationship between QCD4 and QED2, we shall
briefly review and summarize previous QED2 results in order to show
how the Wigner function of produced particles in string fragmentation
can be evaluated.  In conformity with the usual notation, we denote
the coupling constant in QED2 by $e$ which will subsequently take on
values related to the QCD2 coupling constant $-g_{\rm 2D}$ (with the
choice of the negative sign by convention) or phenomenologically as
$\sqrt{\pi}$ times the transverse mass of a detected pion.  For
convenience, we shall label the two-dimensional coordinate and
momentum by ${\underline x}=(x^0,x^1)=(t,x)$ and ${\underline
p}=(p^0,p^1)$, respectively.

From the work of Schwinger \cite{Sch62}, it is known that QED2
involving massless fermions with electromagnetic interactions is
equivalent to a free boson field $\phi$ with a mass $m=e/\sqrt{\pi}$,
where $e$ is the coupling constant.  We can understand this remarkable
property of massless QED2 from the following lines of reasoning.  (For
a simple pedagogical discussion of QED2, see Chapter 6 of
\cite{Won94}.)  One starts with charged fermions occupying the
negative-energy Dirac sea.  If there is a current disturbance $j^\mu$
in some region of space, it will lead to an electromagnetic gauge
field $A^\mu$. This electromagnetic field $A^\mu$ will affect all wave
functions of all particles, including those in the Dirac sea.  The
resultant wave functions generate a charge current $j^\mu$.  This
charge current in turn generates a gauge field $A^\mu$ which interacts
with the fermions to generate the current.  The coupling of the gauge
field $A^\mu$ and the fermion current $j^\mu$ in a self-consistent
manner is a problem of great complexity.  Remarkably, when the gauge
invariance property of the current is properly taken into account, the
generated current $j^\mu$ is related to the the introduced local
electromagnetic disturbance.  This current $j^\mu$ in turn leads to an
electromagnetic field $A^\mu$, through the Maxwell equation. If this
generated electromagnetic field is self-consistently the same as the
electromagnetic field $A^\mu$ which was first introduced, then we find
that $\square A^\mu$ is proportional to $A^\mu$.  The electromagnetic
field satisfies the Klein-Gordon equation appropriate for a boson with
a mass, $\square A^\mu + (e^2/\pi)A^\mu=0$.  In the Lorentz gauge, we
can represent the gauge field $A^\mu$ in terms of the boson field
$\phi$ by $ A^\mu=\epsilon^{\mu \nu}\partial_\nu \phi/m$. QED2
involving interacting massless fermions is therefore equivalent to a
field of free $\phi$ boson field with quanta of mass $m$.

The relation between the fermion and boson field quantities is
\cite{Sch62}
\begin{eqnarray}
\label{jmu}
j^{\mu}=-e\epsilon^{\mu \nu} \partial_{\nu}\phi/\sqrt{\pi},
\end{eqnarray}
where $j^{\mu}$ is the fermion current which can be taken to be a
real quantity, and $\epsilon^{\mu \nu}$ is the antisymmetric tensor
$\epsilon^{01}= -\epsilon_{01}=-1 $.   

In QED2, the string fragmentation process is described as the time
evolution of a fermion separating from an antifermion at a
center-of-mass energy $\sqrt{s}$, with the production of the quanta of
the $\phi$ boson field (or the perturbation of the $A^\mu$ field) at
subsequent times.  It can be formulated as an initial-value problem.
If the current $j^\mu$ arising from the fermions is initially known,
then the subsequent dynamics of the boson field $\phi$ can be inferred
at all times.

The stipulation that the interacting fermion field is exactly
equivalent to a free massive boson field facilitates the solution of
such an initial-value problem.  We can represent the dynamics of the
system through the evolution of a real massive pseudoscalar field
$\phi$, with the most general solution
\begin{eqnarray}
\label{phiwfa}
\phi(x,t)=\frac {1}{  \sqrt{2\pi}} \int d^2p \theta(p^0)
\delta(p^2-m^2) \biggl \lbrack c(p^1)e^{-i{\underline p}\cdot{\underline  x}} +
c^*(p^1)e^{+i{\underline p}\cdot {\underline x}} \biggr \rbrack,
\end{eqnarray}
where ${\underline p}\cdot{\underline x}=p^0t-p^1x$.  The coefficient
$c(p^1)$ and its complex conjugation $c^*(p^1)$ can be obtained from
the initial fermion currents $ j^0(x,t),$ and $j^1(x,t)$ at $t=0$.  By
using Eq. (\ref{jmu}), we therefore have
\begin{eqnarray}
[~\frac {e }{ {\sqrt {\pi}}} \partial_x \phi(x,t)]_{t=0} = j^0(x,0),
  \end{eqnarray}
and 
\begin{eqnarray}
[-\frac{e }{ {\sqrt {\pi}}} \partial_t \phi(x,t)]_{t=0} = j^1(x,0).
\end{eqnarray}
From these initial conditions and Eq. (\ref{phiwfa}), we obtain
\begin{eqnarray}
\label{cp1}
c(p^1)= -\frac{ i\sqrt{\pi}} { e}
\biggl \lbrack \frac{p^0 }{ p^1} \tilde {j^0} (p^1) + \tilde {j^1}(p^1)  
\biggr \rbrack   \end{eqnarray}
and
\begin{eqnarray}
\label{cp2} 
c^*(-p^1)=-\frac{ i\sqrt{\pi}}{ e}
\biggl \lbrack \frac {p^0 }{ p^1} \tilde {j^0} (p^1) - \tilde {j^1}(p^1)  
\biggr \rbrack,   \end{eqnarray}
where $\tilde {j^\mu}(p^1)$ is the Fourier transform of $j^\mu (x,0)$,
\begin{eqnarray} 
\label{j0}
\tilde {j^\mu}(p^1) = \frac{1 }{ \sqrt{2\pi}} \int dx e^{-ip^1x} 
{ j^\mu(x,0) }.    
\end{eqnarray}
With the coefficients $c(p^1)$ and $c^*(p^1)$ thus determined, the
energy $P^0$ of the system can be determined from
\begin{eqnarray}
P^0=
\frac{1 }{ 2} \int 
\biggl \lbrack 
 \left ( \frac { { \partial \phi} (x,t)}{ {\partial t} } \right )^2 
+\left (\frac{ { \partial \phi} (x,t) }{ {\partial x} }\right )^2 
+ m^2 \left ( \phi(x,t)\right )^2 
\biggr \rbrack 
dx. 
\end{eqnarray}
It is useful to separate $\phi(x,t)$ in terms of positive and negative
frequency components,
\begin{eqnarray}
\phi(x,t)= \phi^{(-)}(x,t) + \phi^{(+)}(x,t), 
\end{eqnarray}
where
\begin{eqnarray}
\label{phip}
 \phi^{(-)} (x,t) =
\int \frac{dp^1}{\sqrt{2\pi}2p^0} c(p^1)
e^{-i{\underline p}\cdot {\underline x}}, 
\end{eqnarray}
and
\begin{eqnarray}
\label{phim}
 \phi^{(+)} (x,t) =
\int \frac{dp^1}{\sqrt{2\pi}2p^0} c^* (p^1)
e^{i{\underline p}\cdot {\underline x}}.
\end{eqnarray}
The energy of the system is then
\begin{eqnarray}
\label{p0}
P^0=\int dx
\biggl \lbrack 
         \partial_t \phi^{(+)}(x,t)  \partial_t \phi^{(-)}(x,t) 
+        \partial_x \phi^{(+)}(x,t)  \partial_x \phi^{(-)}(x,t) 
+ m^2      \phi^{(+)} (x,t) \phi^{(-)}(x,t) 
\biggr \rbrack .
\end{eqnarray}
This leads to the energy
\begin{eqnarray}
\label{P00}
P^0=\int dp^1 c(p^1)c^*(p^1)/2,    
\end{eqnarray}
which is clearly a time-independent quantity.  The momentum
distribution of the bosons is given by \cite{Won91}
\begin{eqnarray}
\label{cpcp}
\frac{dN }{ dp^1}
&=&\frac{c(p^1)c^*(p^1)}{2p^0}\nonumber\\
&=&\frac{\pi }{ 2p^0e^2} 
\biggl \lbrack \frac{p^0 }{ p^1} \tilde {j^0} (p^1) + \tilde {j^1}(p^1)  
\biggr \rbrack
\biggl \lbrack \frac{p^0 }{ p^1} \tilde {j^0} (-p^1) + \tilde {j^1}(-p^1)  
\biggr \rbrack,
  \end{eqnarray}
where $p^0=\sqrt{(p^1)^2+m^2}$.  The rapidity distribution of the produced
particles is \cite{Won91}
\begin{eqnarray}
\frac{dN }{ dy}=
\frac{\pi }{ 2e^2} 
\biggl \lbrack \frac {p^0}{ p^1} \tilde {j^0} (p^1) + \tilde {j^1}(p^1)  
\biggr \rbrack
\biggl \lbrack \frac{p^0 }{ p^1} \tilde {j^0} (-p^1) + \tilde {j^1}(-p^1)  
\biggr \rbrack.
  \end{eqnarray}
We have thus obtained a simple relation between the rapidity
distribution and the Fourier transforms of the initial fermion
charge currents.

To obtain the Wigner function of the produced particles, we rewrite
Eq. (\ref{p0}) as
\begin{eqnarray}
\label{p0a}
P^0=\int dx \, dk \int \frac{ds \, e^{iks}}{2\pi}
\biggl \lbrack 
        \partial_t \phi^{(+)}(x+\frac{s}{2})
        \partial_t \phi^{(-)}(x-\frac{s}{2})  
+       \partial_x \phi^{(+)}(x+\frac{s}{2})
        \partial_x \phi^{(-)}(x-\frac{s}{2})  
+ m^2              \phi^{(+)}(x+\frac{s}{2}) 
                   \phi^{(-)}(x-\frac{s}{2}) 
\biggr \rbrack .
\end{eqnarray}
The total energy of the system is related to the Wigner function
$f(x,k)$ by
\begin{eqnarray}
\label{p0b}
P^0=\int dx ~dk ~k^0 f(x,k),
\end{eqnarray}
where $k^0=\sqrt{k^2+m^2}$, 
and the total number $N$ is
\begin{eqnarray}
N=\int dx ~dk f(x,k).
\end{eqnarray}
By comparing Eqs.\ (\ref{p0a}) with (\ref{p0b}), the Wigner function
of produced bosons is given by
\begin{eqnarray}
 f(x,k)=
\int \frac{ds~e^{iks}}{2\pi k^0}
\biggl \lbrack 
        \partial_t \phi^{(+)}(x+\frac{s}{2})
        \partial_t \phi^{(-)}(x-\frac{s}{2})  
+       \partial_x \phi^{(+)}(x+\frac{s}{2})
        \partial_x \phi^{(-)}(x-\frac{s}{2})  
+ m^2              \phi^{(+)}(x+\frac{s}{2}) 
                   \phi^{(-)}(x-\frac{s}{2}) 
\biggr \rbrack .
\end{eqnarray}
Upon substituting the explicit forms of $\phi^{(\pm)}$ in terms of
$c(p^1)$ and $c^*(p^1)$ from Eqs. (\ref{phip}) and (\ref{phim}), we
obtain
\begin{eqnarray}
\label{wig}
 f(x,k)=
\frac{1}{4\pi k^0}
\int dq e^{iq[x-(p^0(q)-{p^{0}}'(q))t]}
\frac{p^0{p^{0}}'+(k^0)^2 - q^2/4}{2 p^0{{p^{0}}'}} 
c(k+\frac{q}{2}) c^*(k-\frac{q}{2}),
\end{eqnarray}
where $p^0=\sqrt{(k+q/2)^2+m^2}$ and ${p^{0}}'=\sqrt{(k-q/2)^2+m^2}$.
We can alternatively rewrite the above as
\begin{eqnarray}
\label{wig1}
 f(x,k)=
\frac{1}{2\pi k^0}
\int_0^\infty
 dq \cos \{q[x-(p^0(q)-{p^{0}}'(q))t]\}
\frac{p^0{p^{0}}'+(k^0)^2 - q^2/4}{2 p^0{p^{0}}'} 
c(k+\frac{q}{2}) c^*(k-\frac{q}{2}).
\end{eqnarray}
Therefore, if an initial fermion current is given, its spatial
Fourier components $\tilde {j^\mu}(p^1)$ can then be evaluated using Eq.\
(\ref{j0}), and the spatial Fourier coefficients $c(p^1)$ and $c^*(p^1)$
of the bosons can be obtained from Eqs.\ (\ref{cp1}) and (\ref{cp2}).
These Fourier coefficients can be used to evaluate the Wigner function
in Eq. (\ref{wig1}).  

As the longitudinal momentum is often represented by the rapidity
variable $y$, with $k^0=m \cosh y$ and $|{\bf k}|=k=m\sinh y$, we can
represent the Wigner function as a function of the longitudinal
coordinate $x$, and the rapidity $y$ as
\begin{eqnarray}
f(x,y)=k^0 f(x,k)\biggr |_{k=m\sinh y}.
\end{eqnarray}

\section{String fragmentation at a finite energy}

As a simple example, we shall apply the above results to particle
production in the high-energy annihilation of $e^+$ with $e^-$, which
falls within the domain of multi-flavor QCD4.  The approximate
consideration of the dynamics as occurring within a flux tube in the
fragmentation of a string, allows one to approximate the
transversely-confined QCD4 to QCD2.  Multi-flavor massless QCD2 in
turn becomes a set of $N_c^2-1$ Abelian QED2-type systems in the large
$N_f$ limit, with the production of the Schwinger QED2-type states.
Because the boson mass for the realistic case of massive multi-flavor
QCD2 with $N_f\sim 3$ is not known, we shall content ourselves with
using phenomenologically the experimental observed transverse mass of
the produced pions as the mass of the produced bosons.  To take into
account the linear dependence of the multiplicity on the number of
flavors, we can consider the fragmentation of a single-flavor string,
with the flavor number ascribed to be the effective charge $\nu =
\sqrt{N_f}$ of the separating charges.

We therefore start with a positive charge with charge $\nu e$
separating from a negative charge $-\nu e$ at a finite energy $\sqrt
{s}$ in the center-of-mass system, to simulate approximately the
dynamics as a quark separating from an antiquark after an
electron-positron annihilation.  At $t=0$ the fermion and the
anti-fermion charges superimposed so that the total charge density of
the system is zero:
\begin{eqnarray}j^0(x,0)=0.   \end{eqnarray}
However, the initial charge current $j^1(x,0)$ is non-zero, and it is
a symmetric function of $x$.  In the case of an infinite energy, the
current can be represented as a sum of two delta functions of the two
sources moving with the speed of light out to $+x$ and $-x$ directions
\cite{Cas74} and the corresponding terms in the boson field $\phi$ are
step functions.  This suggests a simple generalization of the initial
$\delta$-function current for finite energies, in which one replaces
their step-functions in $\theta(x\pm t)$ by a smooth function of
diffusivity $\sigma$ \cite{Won91},
\begin{eqnarray}
\phi(x,t)=-\frac{\sqrt{\pi}\nu}{2} \left ( \tanh\frac{x+t}{\sigma}
                                          -\tanh\frac{x-t}{\sigma} \right ).
\end{eqnarray}
The diffusivity $\sigma$ is related to the total invariant mass $\sqrt
{s}$ of the system given in Eq. (\ref{sigma}) below.  From
Eq. (\ref{jmu}), the field $\phi(x,t)$ leads to the current
$j^\mu(x,t)=0$:
\begin{eqnarray}
j^0(x,t)=\frac{e\nu}{2\sigma} \left
( \frac{1}{\cosh^2\left ( \frac{x+t}{\sigma} \right )} -
\frac{1}{\cosh^2\left ( \frac{x-t}{\sigma} \right )} \right ).
\end{eqnarray}
and
\begin{eqnarray}
j^1(x,t)=\frac{e\nu}{2\sigma} \left
( \frac{1}{\cosh^2\left ( \frac{x+t}{\sigma} \right )} +
\frac{1}{\cosh^2\left ( \frac{x-t}{\sigma} \right )} \right ).
\end{eqnarray}
Strictly speaking, the fermion currents should travel with a speed
slightly less than the speed of light.  At high energies, the
difference between their speed and the speed of light is so small that
it can be neglected.  At $t=0$, we have $j^0(x,0)=0$ and
\begin{eqnarray}j^1(x,0)= \frac{\nu e}{ {\sigma \cosh ^2 ({x/\sigma}}) }.  
\end{eqnarray}
For this initial current distribution $j^\mu(x,0)$ the Fourier
transforms of $j^\mu(x,0)$ are
\begin{eqnarray}
\label{j0t}
\tilde {j^0} (p^1)
=0, 
\end{eqnarray}
and
\begin{eqnarray}
\label{j1t}
\tilde {j^1} (p^1)
=-\frac{iec(p^1) }{ \sqrt{\pi}} 
=\frac{iec^*(p^1) }{ \sqrt{\pi}} 
=\frac{ { \nu e\pi p^1 \sigma} }{ { \sqrt{2\pi} \sinh (\pi p^1 \sigma /2)} }.
\end{eqnarray}
From Eqs. (\ref{cp1})-(\ref{cp2}), these initial charge current
Fourier components leads to the Fourier coefficient
\begin{eqnarray}
\label{newc}
c(p^1)=i\sqrt{2} \frac{\nu p^1\pi\sigma/2}
{\sinh(p^1\pi\sigma/2)}.
\end{eqnarray}
We can use Eq. (\ref{P00}) to calculate the energy $P^0=\sqrt s$ which is
independent of time, we obtain a relation between 
$\sigma$ and $\sqrt{s}$:
\begin{eqnarray}
\label{sigma}
\sigma=\frac{ 2\pi \nu^2 }{ {3\sqrt{s}}}.  \end{eqnarray} 
The rapidity distribution is therefore \cite{Won91}
\begin{eqnarray}
\label{dndy}
\frac{ dN }{ dy } =\frac{ {\nu^2 \xi^2} }{ \sinh^2{\xi} },   
\end{eqnarray}
where
\begin{eqnarray}
\label{xi}
\xi=\frac{ { \nu^2 \pi^2 m \sinh y} }{ {3 \sqrt{s}} }.   
\end{eqnarray}
Thus, the rapidity distribution is boost-invariant, $dN/dy=\nu^2$, in
the limit of very high energies.  At a finite energy $\sqrt{s}$, the
rapidity distribution therefore turns into a plateau structure, with a
half-width $y_0$ at half maximum characterized by $\xi_0=1.491$
and 
\begin{eqnarray}
y_0=\sinh^{-1}\left (\frac{3 \xi_{0} \sqrt{s}}
                                   {\nu^2 \pi^2 m} \right ).
 \end{eqnarray}
The half-width $y_0$ varies with $\sqrt{s}$ of the fragmenting string
as $y_0 \propto \sinh^{-1} (\sqrt{s}) \sim \ln ( \sqrt{s})$, as
expected.

\section{Evaluation of the Wigner function for a separating 
fermion and antifermion pair}

\begin{figure} [h]
\includegraphics[angle=0,scale=0.50]{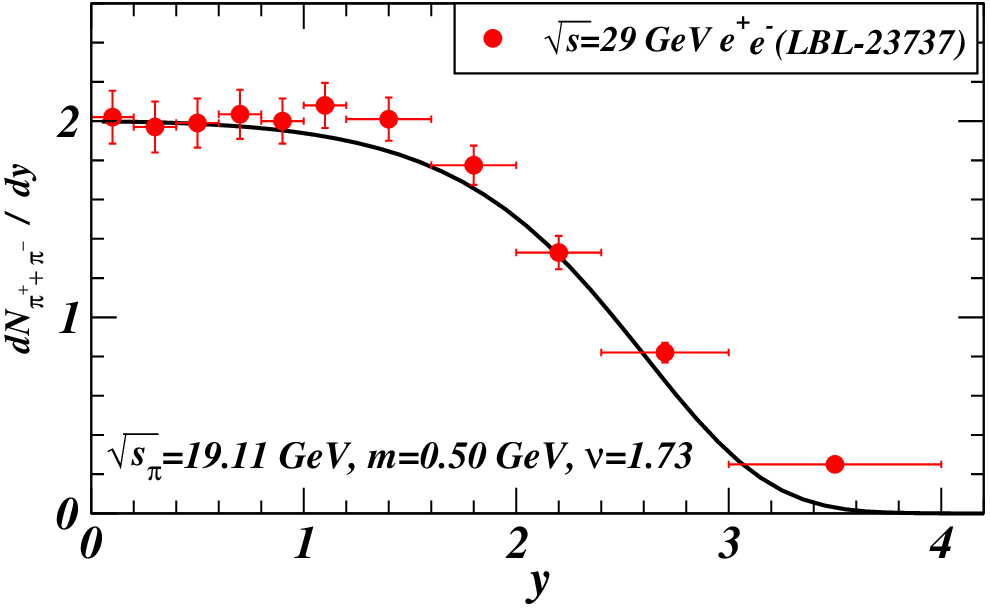}
\vspace*{0.0cm} 
\caption{ (Color online) The rapidity distribution of pions in
  high-energy $e^+$-$e^-$ annihilation at $\sqrt{s}=$29 GeV, out of
  which $\sqrt{s_\pi}$=19.11 GeV goes into producing pions.  The data
  is from \cite{Aih88,Hof88,Aihnote} and the solid curve is from Eq.\
  (\ref{dndy}) with $m=0.50$ GeV, $\nu=1.73$.  }
\end{figure}

\begin{figure} [b]
\includegraphics[angle=0,scale=0.50]{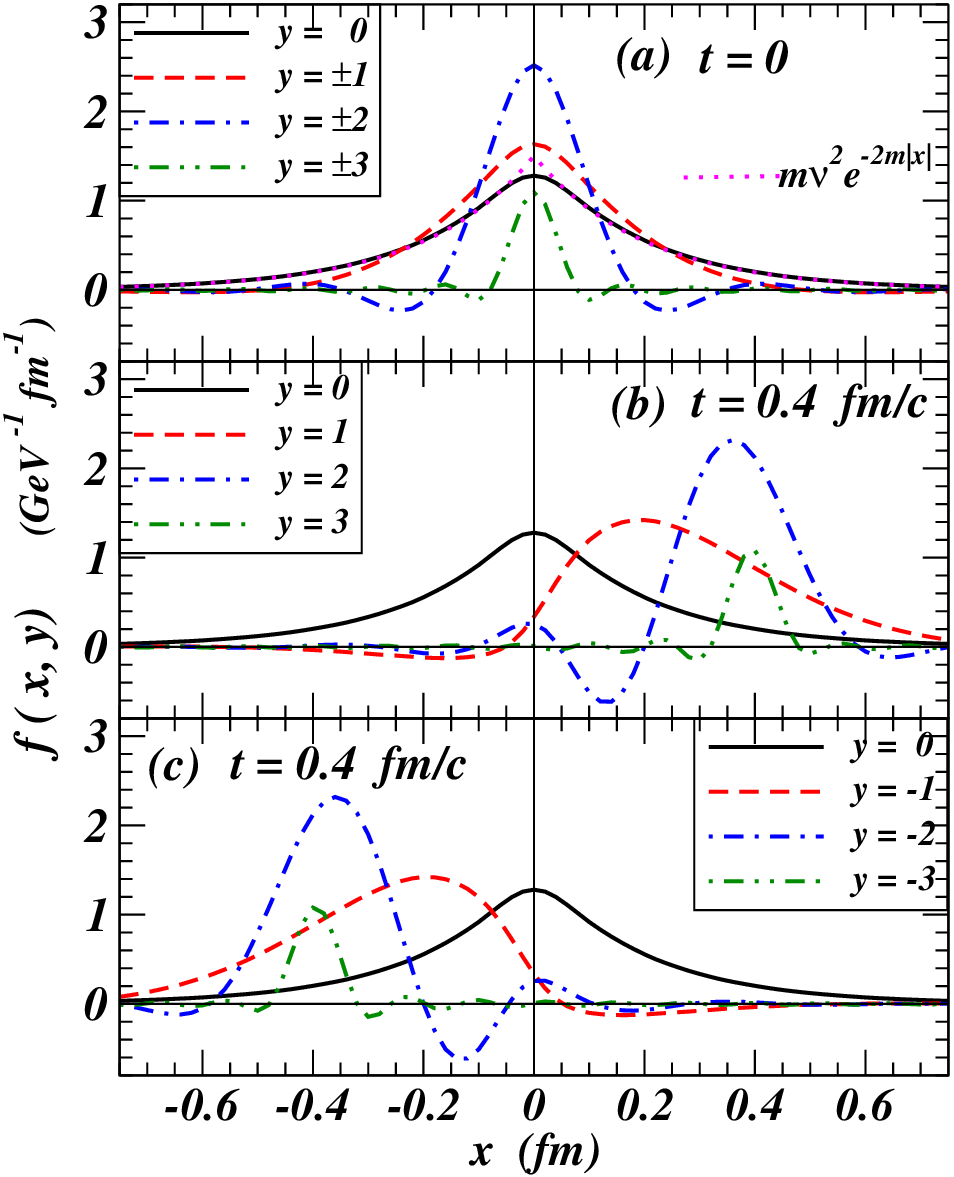}
\vspace*{0.0cm} 
\caption{ (Color online) Wigner function $f(x,y)$ as a function of $x$
for different values of positive $y$.  Fig.\ 2(a) gives $f(x,y)$ for
$t=0$, and the dotted curve is the approximate analytical result for
$f(x,y=0)$ at $t=0$ given by Eq.\ (\ref{vu2}).  Fig.\ 2(b) gives
$f(x,y)$ for positive $y$ at $t=0.4$ fm/c, and Fig.\ 2(c) for negative
$y$ at $t=0.4$ fm/c.  }
\end{figure}

It is of interest to study a specific numerical example to illustrate
the space-time dynamics of string fragmentation.  We can consider the
experimental rapidity distribution of charged pions in $e^+$-$e^-$
annihilation at $\sqrt{s}=29$ GeV \cite{Aih88,Hof88}.  In such an
annihilation, heavier particles such as kaons and baryons are also
produced and they carry a fraction of the initial energy.  By
integrating out the experimental $dN/dy p_t dp_t$ data in Table 11 of
Ref.\ \cite{Aih88}, the particle multiplicities of the produced pions,
kaons, and baryons are in the ratio of $N_\pi: N_K: N_N = 80:15:5$ and
the total energies of the produced particles are 19.11, 7.10, and 2.53
GeV for pions, kaons, and nucleons respectively \cite{Aih88}.

Rather than attempting to model the production of heavier mesons here,
we shall instead simply consider pions only and study an idealized
situation in which the system fragments only into pions. We can
consider the pions to take up $\sqrt{s_\pi}=$19.11 GeV of the total
energy and fit the rapidity distribution of the pions with Eqs.\
(\ref{dndy}) and (\ref{xi}) which contain only two unknown parameters:
the mass $m$ and the effective charge $\nu$.  The pions have an
average transverse momentum of 0.48 GeV \cite{Aih88}, which
corresponds to a pion transverse mass of $m_{\pi
{}_T}=\sqrt{p_T^2+m_\pi^2}=0.50$ GeV.  In approximately compactifying
QCD4 to QCD2 and QED2, we have subsumed the details of the transverse
degree of freedom by using the transverse mass of quarks.  With the
quark mass modified to be the transverse quark mass by this
compactification, the corresponding produced boson mass should also be
modified to be the transverse boson mass of the observed boson.
Accordingly, the boson mass $m$ of the produced particle in Eqs.\
(\ref{xi}) should be taken to be the pion transverse mass of $m=m_{\pi
{}_T}=0.50$ GeV.  The theoretical rapidity distribution of charged
pions can then be calculated with Eq. (\ref{dndy}) with the effective
charge $\nu$ as the only unknown parameter.  The effective charge
value of $\nu=\sqrt{3}$ \cite{Won91} gives a good description of the
experimental rapidity plateau of the produced pions
\cite{Aih88,Hof88,Aihnote}, as shown in Fig. 1.  It is interesting
that as $N_f$ is related to $\nu$ by $N_f=\nu^2$.  It appears
phenomenologically that the effective number of flavors participating
in the excitation of the vacuum for $e^+$-$e^-$ annihilation at this
energy is 3. This effective charge is in line with the discussions in
in Section III, where the strengths of the underlying color source
generating the excitation of the gauge fields is proportional linearly
to the number of flavors $N_f$.

From Eq.\ (\ref{sigma}), the width parameter of the initial charge
current $\sigma$ is 0.33/GeV or 0.065 fm, which is a narrow current
distribution.  With the knowledge of $\sigma$ and $\nu$, the Fourier
coefficient $c(p^1)$ can be evaluated the Wigner function from Eq.\
(\ref{wig1}) by direct numerical integration.  Fig.\ 2 shows the
Wigner function $f(x,y)$ as a function of $x$, for various values of
$y$ at $t=$0 in Fig.\ 2(a), for positive $y$ at $t=$0.4 fm/c in Fig.\
2(b), and for negative $y$ at $t=$0.4 fm/c in Fig.\ 2(c).  The range
of rapidities in Fig.\ (1) span over different regions of the rapidity
plateau as one can see in Fig.\ 1. One notes the following interesting
features.

\begin{enumerate}

\item At $t$=0, the peak of all Wigner functions occur at $x=0$,
  indicating that all produced bosons with different momenta in 
  different regions of the rapidity plateau are present at $t=0$.
  This picture is in contrast to the momentum-space-time ordering of
  the classical picture of particle production where particles of
  larger momenta $|y|$ are produced at larger $|x|$ at a later time.

\item 
The Wigner function is symmetrical with respect to the changes of both
$x$ to $-x$ and $y$ to $-y$.  That is, $f(x,y;t)=f(-x,-y;t)$.

\item 
The Wigner function for the zero-mode with $y=0$ behaves approximately as
$\nu^2 e^{-2m|x|}$ (see Eq.\ (\ref{vu2}) below) and is positive
definite.  The width of the dominant Wigner function peak for
$|y|\ne 0$ decreases as a function of increasing $|y|$, as expected
from the uncertainty principle.

\item Except for Wigner function of the zero-mode with $y=0$ which
  is positive definite, the Wigner function of other longitudinal
  momenta oscillates as a function of space and time and can assume
  negative values at locations away from the dominant peaks.

\item Except for Wigner function of the zero-mode with $y=0$ which
  does not depend on time, the dominant peaks of the Wigner function
  moves to the positive longitudinal direction for positive $y$ and to
  the negative longitudinal direction for negative $y$.  The speed of
  the movement of the peak position increases with an increasing
  magnitude of the longitudinal momentum.

\end{enumerate}
From the above features, one can readily understand the movement of
the peak positions of the Wigner functions for different $y$ values, as
they follow the motion of what is expected from classical physics.

How do we understand the relatively large width of the spatial
distribution of the Wigner function of the zero-mode with $y=0$?  The
half width of the distribution is about 0.2 fm, which is large
compared to the width parameter $\sigma \sim$ 0.065 fm for the charge
current.  What physical quantity determines the scale of this width?
It is instructive to check the Wigner function for this zero mode
case.  For $y=0$, the Wigner function becomes
\begin{eqnarray}
f(x,y=0)&=&\frac{1}{2\pi}\int_0^\infty dq\, \cos(qx) \frac{m^2}{(q^2/4)+m^2}
2\nu^2 \frac{(q\pi\sigma/4)^2}{\sinh^2(q\pi\sigma/4)},
\nonumber\\
&\sim & 
\frac{\nu^2}{\pi}\int_0^\infty dq\, \cos(qx) \frac{m^2}{(q^2/4)+m^2}
\Theta(\Omega -|q|),
\end{eqnarray}
where $\Omega\sim 2\xi_0/\pi \sigma=2\times1.491/\pi\sigma$.  In the
spatial region where $\Omega x \gg 1$, we have
\begin{eqnarray}
\label{vu2}
f(x,y=0)& \sim &\frac{4xm^2\nu^2}{\pi}\int_0^{\Omega x} d\xi\, \frac{\cos \xi}{ 
{\xi ^2+4m^2x^2}},
\nonumber\\
&  \sim   & m \nu^2 e^{-2m|x|}. 
\end{eqnarray}
We plot the function $m \nu^2 e^{-2m|x|}$ as the dotted curve in Fig.\
2(a).  We find that except in the region of $x \lesssim 0.05$ fm, it
agrees with the exact Wigner function $f(x,y=0)$.  It is also
independent of $t$, as $p^0-{p^0}'=0$ for $k=0$ in Eq.\
(\ref{wig1}). The analytical result shows that the width of Wigner
function for the zero mode with $y=0$ is governed by the mass boson
mass as (width)$\sim \hbar/m$.

How do we understand the origin of the large range of momentum $y$
values that are present in the Wigner function at $t=0$?  From our
derivation of the Wigner function, we note that the presence of an
initial charge current with a very narrow width in $j^\mu(x,t)$ leads
to a wave packet of the boson field $\phi$ in Eq.\ (\ref{phiwfa}), as
a coherent sum of a large number of waves of different momentum
components $p^1$.  The greater the energy of the fragmenting string,
the narrower is the initial spatial current $j^\mu(x,t)$, and the
greater will be the spread of the momentum components of the boson
wave packet $\phi(x,t)$, as indicated by the large width of
$|c(p^1)|$.  The coherent sum of the boson field gives rise to the
corresponding Wigner function that depends on the correlation between
the Fourier components at different momenta.  The Wigner function has
a spatial extension much greater than that of the initial charge
current.

In physical terms, there are two peculiar quantum effects that make
QED2 different from those from classical considerations of string
fragmentation. Firstly, the produced bosons occur initially in a
spatial region more extended than the width of the initial fermion
current. They are governed more by the produced boson mass and the
uncertainty principle than by the spatial width of the initial
current.  Secondly, produced particles of different momenta emerge at
the initial moment of string fragmentation. These peculiarities does
not violate causality because the dynamics of string fragmentation in
QED2 is not just a two-body problem involving only the fragmenting
valence fermion and antifermion. It involves a many-body problem of
particles, antiparticles, and gauge fields interacting
self-consistently with particles in the Dirac sea over an extended
region of space and time, giving rise to the excitations of the vacuum
which manifests themselves as quanta of the massive boson field
\cite{Sch62,Ili84,Won94}.

\section{Interference in the fragmentation of many identical strings}

The process of string fragmentation is a quantum phenomenon. We
therefore expects effects of quantum interference when many identical
strings fragment.  As a well-defined interference phenomenon, it
possesses its own intrinsic theoretical interest.  Furthermore, the
presence or absence of the characteristics of interference may be used
to provide useful information on the fragmenting strings concerning
their properties of being identical or non-identical, when they occur
in close vicinity of each others.

We consider the occurrence of $n$ identical strings at longitudinal
locations $\chi_i$ centered around the longitudinal origin, in the
center-of-mass frame. For simplicity, we shall take the fragmentation
to be occurring at the same time and that the strings are identical in
their characteristics.  In this case, the fragmentation can be
idealized as $n$ interacting fermion-antifermion pairs.  In each pair,
the fermion and the antifermion pulls apart in opposite directions.
The time-like fermion current $j^0(x,0)=0$ is initially zero at all
spatial points and the initial space-like current $j^1(x,0)$ pulls
each fermion-antifermion pair away from each other.

In representing the sum of the fermion currents in a many-string
system, we would like to introduce the concept of the sense $\zeta$ of
a fragmenting string. Consider the fragmentation of a string starting
with a fermion and an antifermion locating at a spatial point.  The
sense of a string specifies the direction of motion of the fermion,
with the antifermion traveling in the opposite direction.  We choose
the convention that the sense $\zeta$ is equal to 1 when the fermion
travels to the positive longitudinal direction and $\zeta$ is -1 when
the fermion travels to the negative longitudinal direction.  The sense
of motion clearly is immaterial for a single string.  However, they
are important in the fragmentation of a many-string system, as we
shall see.

The total current $j_{\rm tot}^\mu(x,t)$ from the $n$
fermion-antifermion pairs can be described as
\begin{eqnarray}
j_{\rm tot}^0(x,t)=\frac{e\nu}{2\sigma} \sum_{i=1}^n \zeta_i \left (
\frac{1}{\cosh^2\left ( \frac{x-\chi_i +t}{\sigma} \right )} -
\frac{1}{\cosh^2\left ( \frac{x-\chi_i-t}{\sigma} \right )} \right ),
\end{eqnarray}
and
\begin{eqnarray}
j_{\rm tot}^1(x,t)=\frac{e\nu}{2\sigma}  \sum_{i=1}^n \zeta_i \left
( \frac{1}{\cosh^2\left ( \frac{x-\chi_i +t}{\sigma} \right )} +
\frac{1}{\cosh^2\left ( \frac{x-\chi_i -t}{\sigma} \right )} \right ),
\end{eqnarray} 
where the width parameter $\sigma$ will need to be self-consistently
determined in terms of the total energy $\sqrt{s}$ of the $n$-string
system because the strings will interfere with each other in
generating the particle rapidity spectrum.

The Fourier transform of the initial current can be easily obtained
and they are related to ${\tilde j}^\mu(p^1)$ of Eqs. (\ref{j0t}) and
(\ref{j1t}) by
\begin{eqnarray}
{\tilde j}_{\rm tot}^\mu (p^1)
=\sum_{i=1}^n \zeta_i e^{-ip^1\chi_i} {\tilde {j^\mu}} (p^1).
\end{eqnarray}
Similarly, the total Fourier component for Eq.\ (\ref{cp1}) is
\begin{eqnarray}
c_{\rm tot} (p^1)
=\sum_{i=1}^n \zeta_i e^{-ip^1\chi_i} c (p^1),
\end{eqnarray}
where $c(p^1)$ is given by Eq.\ (\ref{cp1}).  As a consequence,
Eq.\ (\ref{cpcp}) for the momentum distribution of produced particles
is modified to be
\begin{eqnarray}
\frac{dN_{\rm tot} }{ dp^1} &=&\frac{c_{\rm tot}(p^1)c_{\rm
    tot}^*(p^1)}{2p^0}\nonumber\\ &=&\left [ n + \sum_{i>j}^n 2 
\zeta_i \zeta_j \cos\{
  p^1(\chi_i-\chi_j)\} \right ] \frac{c(p^1)c^*(p^1)}{2p^0}.
\end{eqnarray}
The rapidity distribution for the fragmentation of $n$ strings is
\begin{eqnarray}
\frac{dN_{\rm tot} }{ dy}
=\left [ n + \sum_{i>j}^n 2 \zeta_i \zeta_j 
\cos\{ (m\, \sinh y) (\chi_i-\chi_j)\} \right ] 
\frac{c(p^1)c^*(p^1)}{2}.
\end{eqnarray}
For the Fourier coefficient  given by Eqs.\ (\ref{newc}), we obtain
\begin{eqnarray}
\frac {dN_{\rm tot} }{ dy}
=\left [ n + \sum_{i>j}^n 2 \zeta_i \zeta_j 
\cos\{ (m\, \sinh y\,) (\chi_i-\chi_j)\} \right ] 
\frac{\nu^2 \xi^2}{\sinh^2 \xi},
\end{eqnarray}
where $\xi={ \nu^2 \pi^2 m \sinh y}/ {3 \sqrt{s}} $ is given by Eq.\
(\ref{xi}).  This equation can also be written as
\begin{eqnarray}
\frac{dN_{\rm tot} }{ dy}
=\left [ n + \sum_{i>j}^n 2 \zeta_i \zeta_j \cos\{ (m\, \sinh y)\, (\chi_i-\chi_j)\} \right ] 
\frac {dN_{\rm single} }{ dy},
\end{eqnarray}
where $dN_{\rm single}/dy= \nu^2 {\xi^2}/{\sinh^2 \xi}$. The total energy
of the $n$-string system is given by
\begin{eqnarray}
\label{tots}
\sqrt{s} =\int dy\, m\, \zeta_i \zeta_j \cosh y\, \left [ n + \sum_{i>j}^n 2 \zeta_i \zeta_j \cos\{
  (m \, \sinh y) \, (\chi_i-\chi_j)\} \right ] \frac{\nu^2 \xi^2}{\sinh^2 \xi},
\end{eqnarray}
which provides a relation between $\sqrt{s}$ and $\sigma$.

The Wigner function for the fragmentation of $n$ strings is now
modified to be
\begin{eqnarray}
 f_{\rm tot}(x,k)&=&
\frac{1}{4\pi k^0}
\int dq e^{iq[x-(p^0(q)-{p^{0}}'(q))t]}
\frac{p^0{p^{0}}'+(k^0)^2 - q^2/4}{2 p^0{{p^{0}}'}} 
\nonumber\\
&\times&
\sum_{i,j=1}^n \zeta_i \zeta_j e^{-i(k+q/2)\chi_i +i(k-q/2)\chi_j} 
c(k+\frac{q}{2}) c^*(k-\frac{q}{2}).
\end{eqnarray}
We can simplify the above expression by noting that
\begin{eqnarray}
e^{iqx} e^{-i(k+q/2)\chi_i +i(k-q/2)\chi_j} 
=e^{iq(x-{\bar \chi}_{ij})-ik(\chi_i - \chi_j)},
\end{eqnarray}
where ${\bar \chi}_{ij}$ is the average of the initial longitudinal
positions of strings $i$ and $j$,
\begin{eqnarray}
{\bar \chi}_{ij}=(\chi_i+\chi_j)/2.
\end{eqnarray}
Therefore, we get
\begin{eqnarray}
\sum_{i,j=1}^n  \zeta_i \zeta_j e^{iqx} e^{-i(k+q/2)\chi_i +i(k-q/2)\chi_j} 
=\sum_i^n e^{iq(x-{ \chi}_{i})}+ \sum_{i>j}^n \zeta_i \zeta_j 
e^{iq(x-{\bar \chi}_{ij})}
2 \zeta_i \zeta_j \cos\{k(\chi_i - \chi_j)\}.
\end{eqnarray}
We can thus write the total Wigner function of the produced particles
in the fragmentation of $n$ strings as
\begin{eqnarray}
\label{mwig}
f_{\rm tot}(x,k)=\sum_{i=1}^n f_{\rm single}(x-\chi_i,k) +\sum_{i>j}^n 
2 \zeta_i \zeta_j \cos \{k(\chi_i-\chi_j)\}f_{\rm single}(x-{\bar \chi}_{ij} ,k).
\end{eqnarray}
The corresponding Wigner function $f_{\rm tot}(x,y)$ can then be
obtained as $f_{\rm tot}(x,y)=k^0 f_{\rm tot} (x,k)$ evaluated at
$k=m \sinh(y)$.

We note from these results that there are important interference
effects between identical strings.  The interference appears in the
form of a cosine function $\zeta_i \zeta_j \cos\{k(\chi_i-\chi_j)\}$
that depends on the momentum of the produced particle, the initial
spatial separation between the strings, and the senses of the strings.
The dependence on the string separation is similar to the interference
in the emission of particles from different sources in intensity
interferometry \cite{Won04}.

In the case when the strings are far apart such that $|\chi_i-\chi_j|$
is large, the cross terms in the Wigner function with $i\ne j$ in
Eq.\ (\ref{mwig}) oscillate rapidly about zero (for non-zero modes).
The cross terms will provide only small average contributions and the
direct terms dominate.  In that case when strings are far apart, $
f_{\rm tot}(x,k) \approx \sum_{i}^n f_{\rm single}(x-\chi_i,k)$ for
$|k|\ne 0$, which is the sum of independent Wigner functions for
separated strings.

In the other extreme, all strings are located at the same spatial
point. Among these $n$ strings, there are $n_+$ strings with the
$\zeta=1$ sense and $n_-=n-n_+$ strings with the opposite sense,
$\zeta=-1$.  Then, when all the strings occur at the same point, we
have
\begin{eqnarray}
dN_{\rm tot}/dy=(n_+-n_-)^2 dN_{\rm single}/dy,
\end{eqnarray}
and
\begin{eqnarray}
f_{\rm tot}(x,k {\rm~or~} y )=(n_+ - n_- )^2 f_{\rm single}(x,k {\rm~or~}y).\\
\end{eqnarray}
If all the strings are aligned in the same direction then $dN_{\rm
tot}/dy=n^2 dN_{\rm single}/dy$.  On the other hand, if $n_+ = n_-$,
then $dN_{\rm single}/dy=0$ and $f_{\rm tot}(x,y)=0$.  In general, the
greater the difference $|n_+-n_-|$, the larger is $dN_{\rm tot}/dy$
and $f_{\rm tot}(x,y)$.  The strength of the Wigner function of one
string of one sense is canceled by the Wigner function of the string
of the opposite sense.  This cancellation of the string strength
cannot be complete if the number of strings is odd.

\section{Wigner function in the fragmentation of identical strings}
\begin{figure} [h]
\includegraphics[angle=0,scale=0.50]{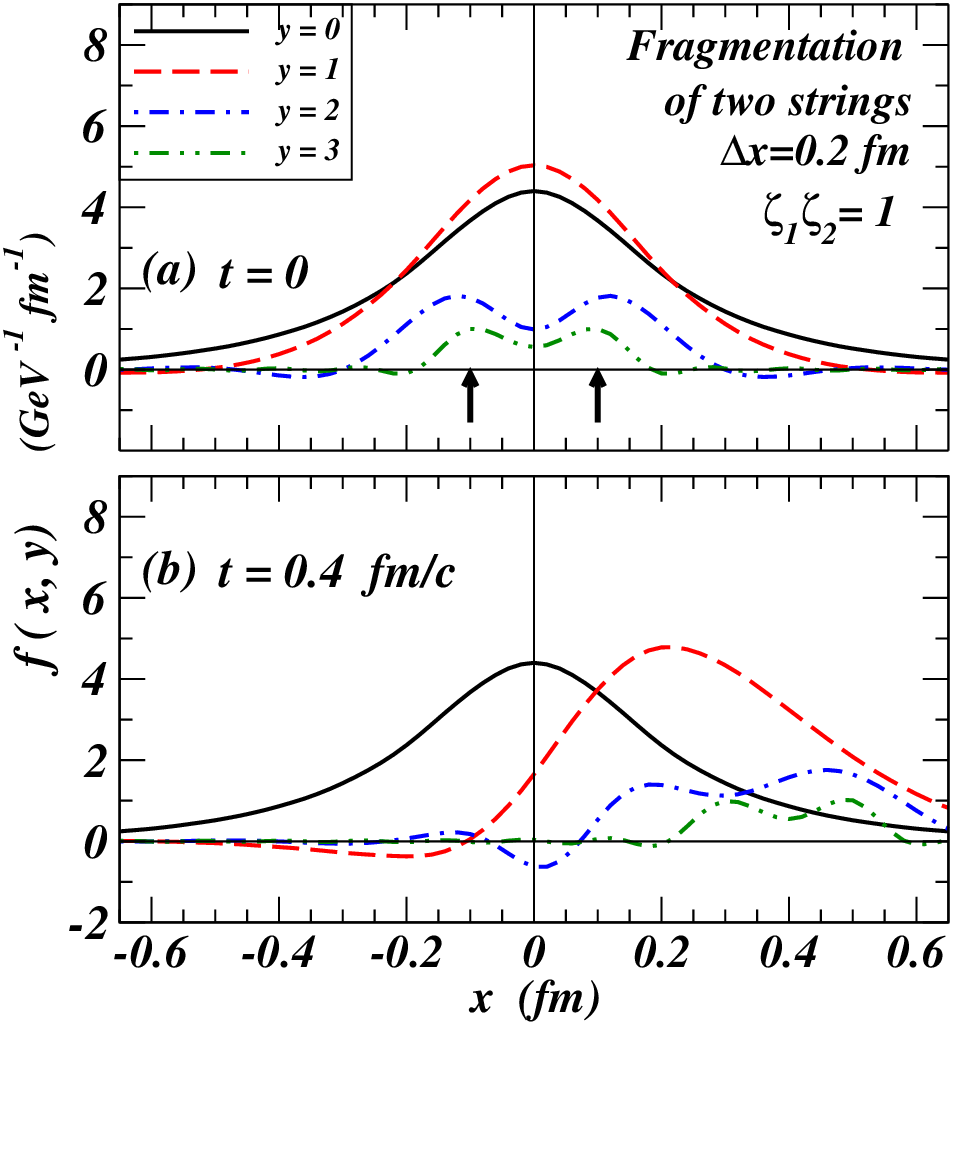}
\vspace*{-1.0cm}
\caption{ (Color online) The Wigner function $f(x,y)$ for two strings
with the same sense, $\zeta_1 \zeta_2=1$.  The strings are initially
located at $x=-0.1$ and $x=0.1$ fm indicated by the two arrows, with a
separation of $\Delta x=0.2$ fm.  Fig.\ 3(a) is for $t=0$ and Fig.\
3(b) is for $t=0.4$ fm/c. }
\end{figure}
As an example, it is instructive to evaluate the Wigner function for a
few cases to see its dependence on the separation between the strings
and their relative sense of directions.  We show the time variation of
the Wigner function for two strings each of which, if separated, has
characteristics the same as those in Fig.\ 2.  We examine first the
case of two identical strings with the same sense so that
$\zeta_1\zeta_2=1$.  In Fig.\ 3, the origins of these two strings are
initially located at $\chi_1=-0.1$ and $\chi_2=0.1$ fm with a
separation $\Delta x=0.2$ fm.  In Fig.\ 4, they are located at
$\chi_1=-0.04$ and $\chi_2=0.04$ fm with a separation $\Delta x=0.08$
fm.  The locations of the origins of the strings are indicated by two
thick arrows in Figs.\ 3 and 4.

\begin{figure} [t]
\includegraphics[angle=0,scale=0.50]{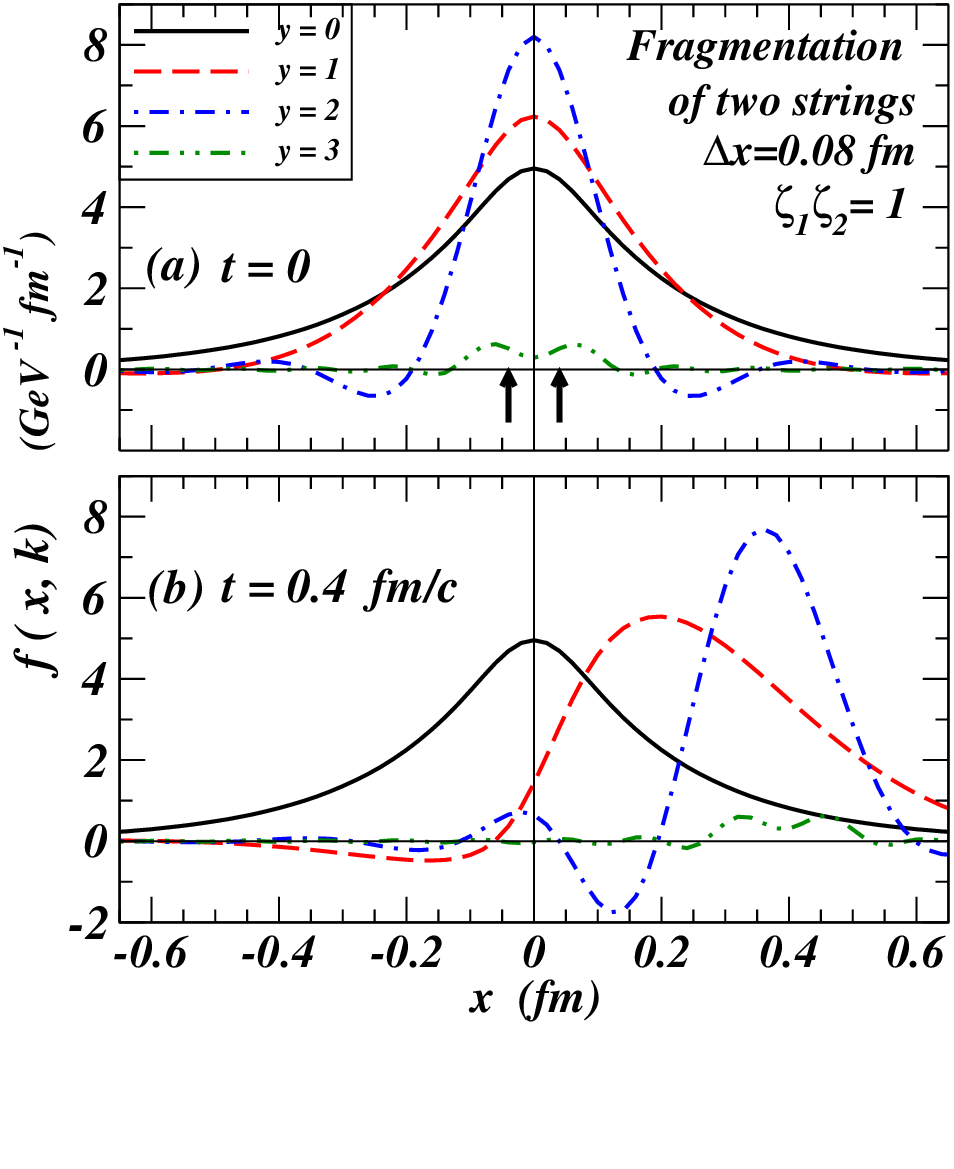}
\vspace*{-1.0cm} 
\caption{ (Color online) Wigner function $f(x,y)$ as a function of $x$
  for different values of $y$ for two strings located at
  $x=-0.04$ and $x=0.04$ fm, as indicated by the two arrows, with a
  separation of $\Delta x=0.08$ fm.   Fig.\ 4(a)
  is for $t=0$ and Fig.\ 4(b) is for $t=0.4$ fm/c. }
\end{figure}

We note the following features from the dynamics of the Wigner
function from Figs. 3 and 4.

\begin{enumerate}

\item
Similar to the case of a single string, all bosons of different
momenta in different regions of the rapidity plateau are present
at $t=0$. This picture is in contrast to the momentum-space-time
ordering of the classical picture of particle production.

\item
The magnitude and the width of Wigner function of the zero-mode with
$y=0$ are quite large when $\zeta_1 \zeta_2=1$.  The magnitude of the
peak is about three times that of Fig.\ 2.  For this case of two
strings, there are three contributions, corresponding to two
contributions centered at the origins of the strings, plus an
additional contribution at $x=0$, as given by Eq.\ (\ref{mwig}).  For
the $y$=0 zero mode, all these three pieces merge together to form a
broad peak at $x=0$ that is independent of time.

\item
For the low rapidity case with $y=1$, the different contributions to
the Wigner function merge into a single peak for both separations of
$\Delta x=0.2$ and 0.08 fm.

\item
For higher rapidities with $y\ge 2$, the different contributions to
the Wigner function merge into a single peak for the case of $\Delta
= 0.08$ fm, but splits into two peaks for $\Delta x=0.2$ fm.  This
shows that the Wigner functions for many strings tend to merge
together into a single structure, when the separation between strings
is small.

\item
As a function of time, the peaks of the Wigner function move in the
positive $x$ direction for positive values of $y$.  They move to
the negative $x$ direction for negative values of $y$.  The low-momentum
Wigner functions with $y=1$ remain to be merged as they
propagate.  For the case of $y\ge 2$ and $\Delta x =0.2$ fm,
there appear two propagating peaks corresponding to the propagation of
the produced particles from the two different strings.  For the case
$y\ge 2$ and $\Delta x = 0.08$ fm, the Wigner functions from the
two strings remain merged during their propagation.

\end{enumerate}

\begin{figure} [t]
\includegraphics[angle=0,scale=0.50]{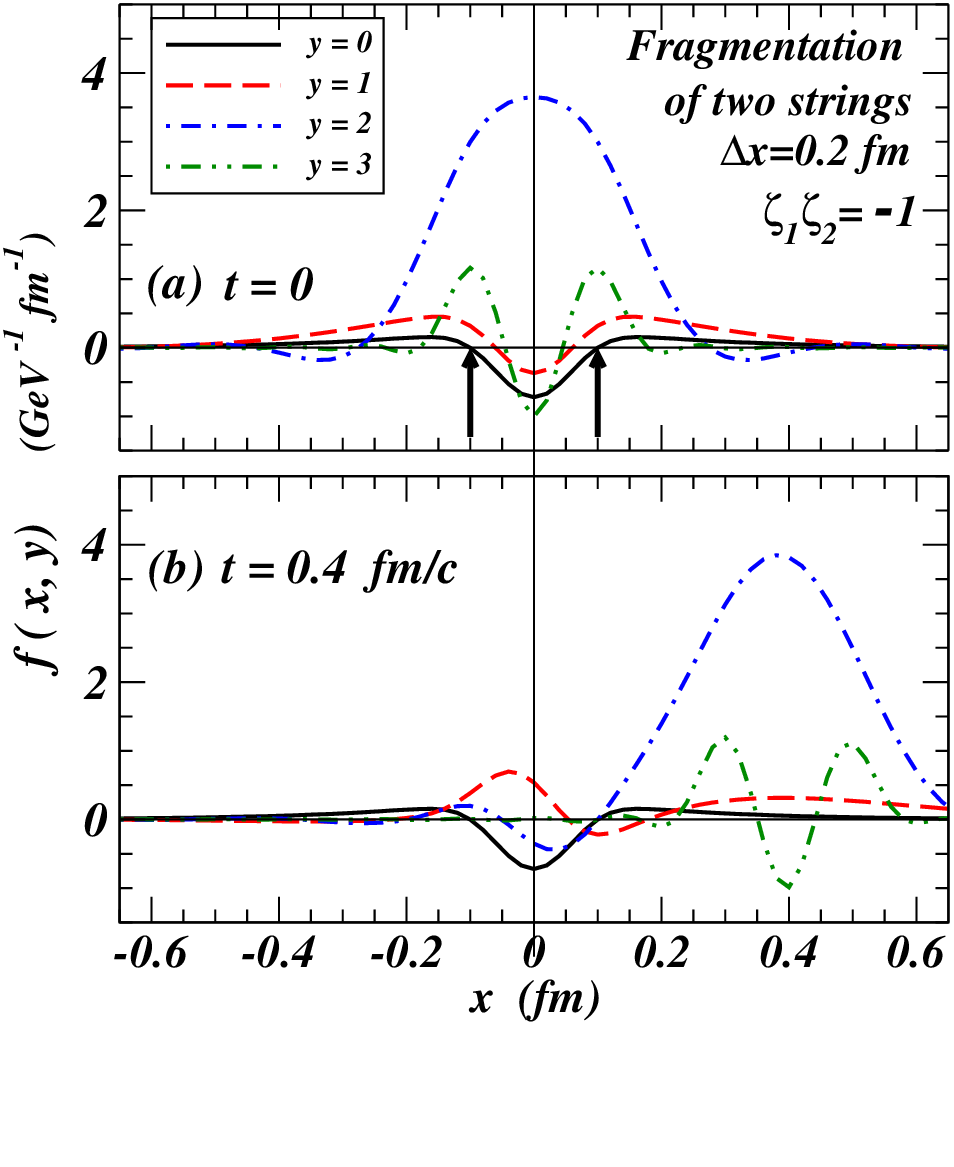}
\vspace*{-1.0cm} 
\caption{ (Color online) The Wigner function $f(x,y)$ as a function of
  $x$ for different values of $y$ for two strings with opposite
  senses, $\zeta_1 \zeta_2=-1$.  The strings are initially located at
  $x=-0.1$ and $x=0.1$ fm indicated by the two arrows, with a
  separation of $\Delta x=0.2$ fm.  Fig. 5(a) is for $t=0$ and
  Fig.\ 5(b) is for $t=0.4$ fm/c. }
\end{figure}

\begin{figure} [p]
\includegraphics[angle=0,scale=0.50]{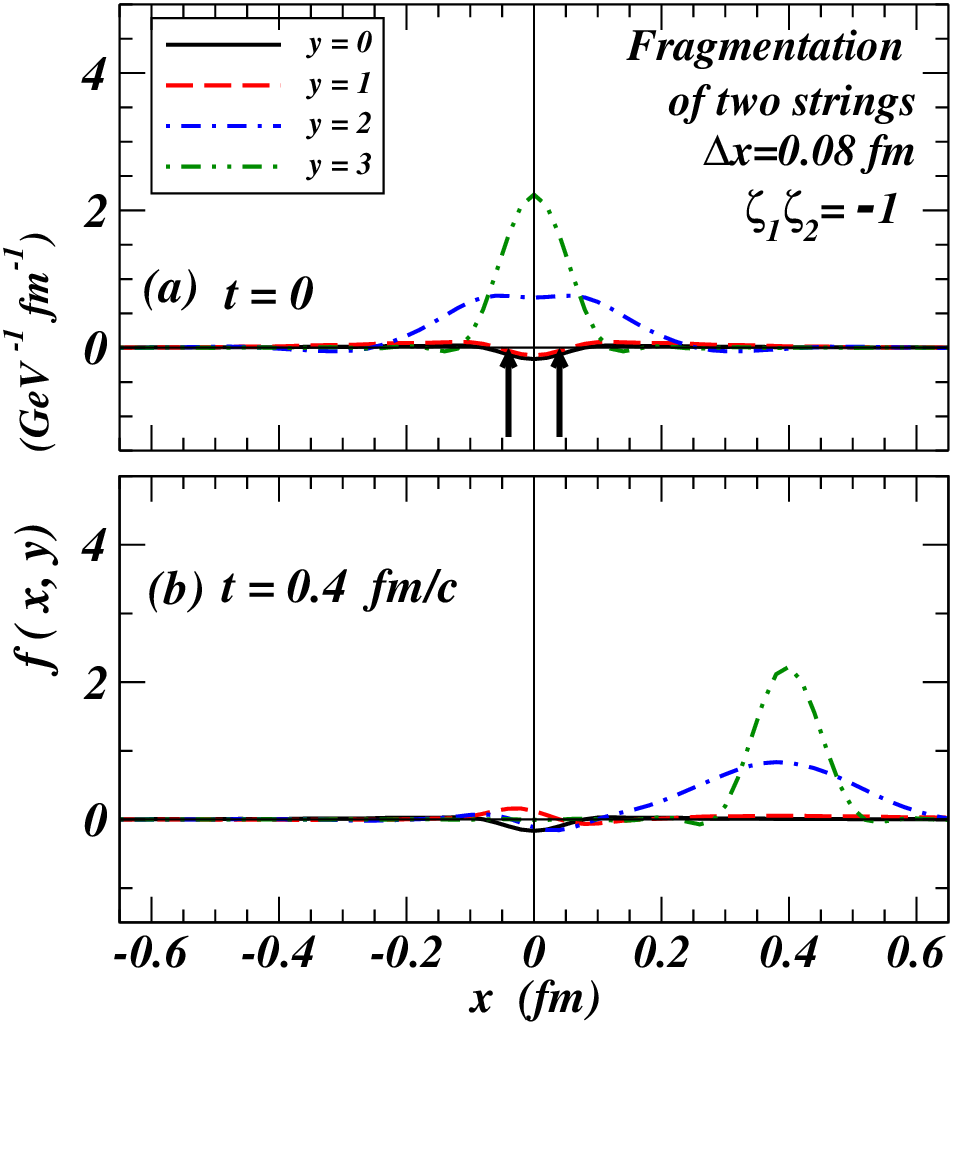}
\vspace*{-1.0cm} 
\caption{ (Color online) The Wigner function $f(x,y)$ as a function of
  $x$ for different values of $y$ for two strings with
  opposite senses, $\zeta_1 \zeta_2=-1$.  The strings are initially
  located at $x=-0.04$ and $x=0.04$ fm indicated by the two arrows, with
  a separation of $\Delta x=0.08$ fm.  Fig. 8(a) is for $t=0$ and Fig.\
  8(b) is for $t=0.4$ fm/c. }
\end{figure}
It is instructive to evaluate next the Wigner function for the case of
two identical strings with the opposite sense so that
$\zeta_1\zeta_2=-1$.  We consider again two strings each of which, if
separated, has characteristics same as those in Fig.\ 2.  In Fig.\ 5,
these two strings are initially located at $\chi_1=-0.1$ and $\chi_2=0.1$
fm with a separation $\Delta x=0.2$ fm.  In Fig.\ 6, they are located
at $\chi_1=-0.04$ and $\chi_2=0.04$ fm, with a separation $\Delta
x=0.08$ fm.  The positions of the origins of the strings are indicated
by the two thick arrows in Figs.\ 5 and 6.

The results in Figs.\ 5 and 6 show that in the fragmentation of two
strings with opposite senses, the magnitude of the Wigner function is
much reduced from the fragmentation of two strings with the same
sense.  The zero-mode Wigner function is negative at $x=0$, in
contrast to the positive definite property of the Wigner function for
a single string or for two strings with the same sense.  There are
oscillations of the Wigner function for different momenta at $t=0$.
For $t>0$, the dominant peaks of the Wigner function propagate
according to their momenta.

As the separation between the strings becomes smaller, the Wigner
function of opposite senses tend to cancel and the magnitude of the
Wigner function decreases, as shown in Fig.\ 6. As we remarked
earlier, the Wigner function is reduced to zero in the limit when the
two identical strings with opposite senses coincide.  

\section{ Dynamics of the Wigner function in the fragmentation of independent 
strings}

\begin{figure} [h]
\includegraphics[angle=0,scale=0.50]{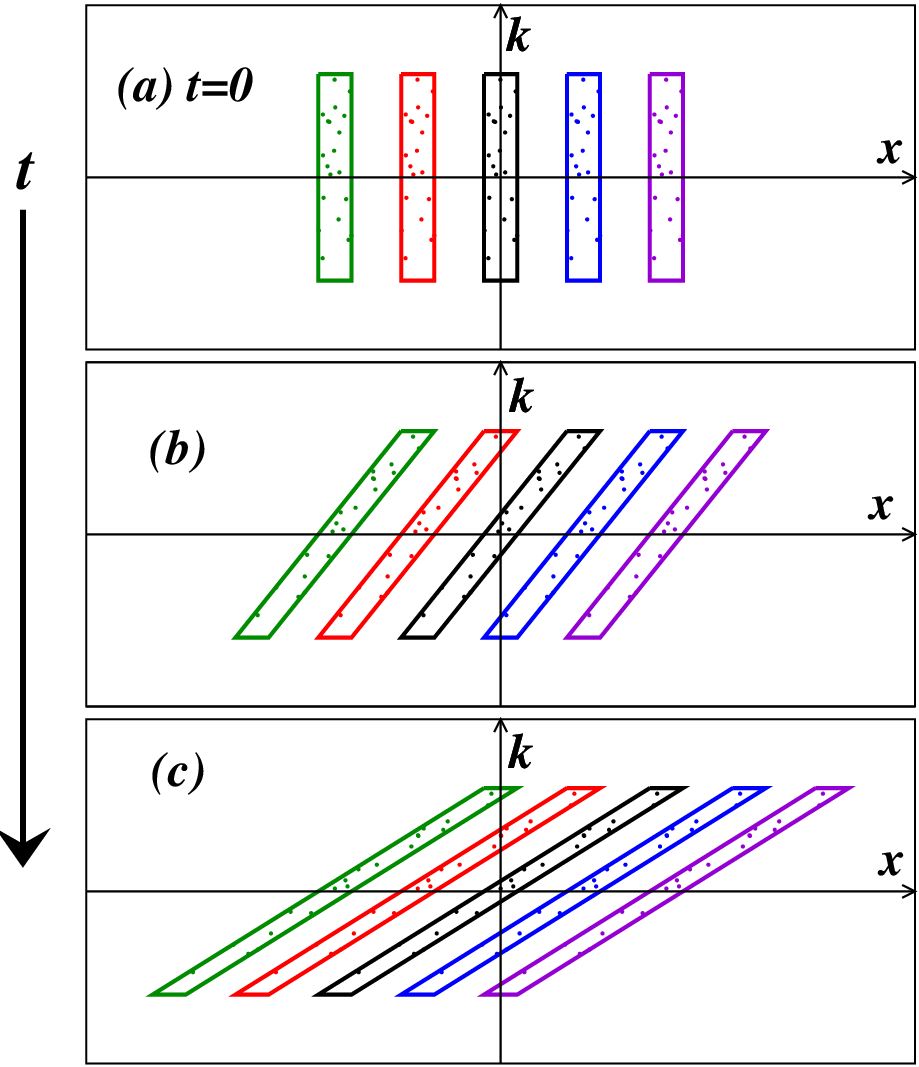}
\vspace*{0.0cm} 
\caption{ (Color online) Schematic space-time evolution of the Wigner
function $f(x,y)$ of produced particles from many independent strings.
Fig.\ 7(a) is for $t=0$, 7(b) and 7(c) are the schematic
configurations at subsequent times.  }
\end{figure}
The last two sections deal with the fragmentation of identical strings
for which interference effects are present.  In a nucleus-nucleus
collision, the number of nucleon-nucleon collisions is large in a
local neighborhood.  We expect that each nucleon-nucleon collision
produces at least two strings spanned between the quark of one nucleon
and the diquark of the other nucleon.  The types of strings that are
formed are however very large in numbers as each string will be
characterized by the color and the flavor of the quark and the diquark
at its two ends, and by the color gauge field component $a$ of
$A^{\nu a}$ the string can excite.  As a consequence, the probability
of a neighboring string to be identical is of order $1/[N_c^2N_f({\rm
q}) N_f({\rm qq})(N_c^2-1)]$ which is quite small.  The production
process in string fragmentation in a nucleus-nucleus collision is
likely to involve the fragmentation of independent non-identical
strings.  It is appropriate to examine the space-time dynamics of
produced particles in the fragmentation of many independent strings.

In a nucleus-nucleus collision in the center-of-mass system, the
average longitudinal separation between neighboring strings is
$d/\gamma$ where $d\sim 1.9$ fm is the average nucleon-nucleon
separation for a nucleus at rest and $\gamma=\sqrt{s_{{}_{NN}}}/m_N$
is the Lorentz contraction factor.  On the other hand, the width of
the Wigner function for the zero mode with $y$=0 is of order
$\hbar/m_{\pi T}$, as indicated by Eq.\ (\ref{vu2}) .  Whether the
Wigner function will be separated or will overlap depend on whether
$d/\gamma$ is greater or less than $\hbar/m_{\pi T}$:
\begin{eqnarray}
\begin{cases}
\label{cri}
{\rm If~~} d/\gamma &\gtrsim  ~~~\hbar/m_{\pi T},
{\rm ~~~strings~are~separated.}\\ 
{\rm If~~} d/\gamma &\lesssim ~~~\hbar/m_{\pi T},
{\rm ~~~strings~overlap.}
\end{cases}
\end{eqnarray}
As the pion transverse mass $m_{\pi T}$ is of order 0.5 GeV, the
Wigner functions will be separated from each other if the collision
energy per nucleon in the center-of-mass system is $\sqrt{s_{{}_{NN}}}
\lesssim 5$ GeV and will overlap each other if $\sqrt{s_{{}_{NN}}}
\gtrsim 5$ GeV.

For each string, particles of different rapidities in different
regions of the rapidity plateau are produced at the moment of string
fragmentation.  They propagate toward opposite longitudinal directions
according to their rapidities.  We can depict the dynamics of the
density of Wigner functions as those of one of the rectangular regions
in Fig.\ 7.

How the partons evolve after production in a heavy-ion collision at
high energies is a complex problem which will require extensive future
investigations.  Our experience with the phase space dynamics of
nucleons in nucleus-nucleus collisions as investigated earlier in
\cite{Won82a,Won82} may furnish an approximate guide to speculate on the
behavior of partons in high energy nuclear collisions. 

We can discuss first the case of $\sqrt{s_{{}_{NN}}} \lesssim 5$ GeV
in the fragmentation of independent and separated strings as shown in
Fig.\ 7.  Produced particles in these separated strings will stream
from one point to another point as time proceeds.  Their streaming
will lead to three important effects. First, they will replenish those
particles that have just streamed away from the point of production,
as is depicted in Fig. 7(b) and 7(c).  Secondly, as a result of their
streaming, they will encounter at the same longitudinal point produced
particles streaming from neighboring points in the opposite direction.
Such encountering will lead to collisions.  The discussion of the
collisions of the produced particles necessitates the inclusion of the
transverse degrees of freedom.  It is necessary to restore the
transverse momentum distribution of the produced particles by using
information on the transverse eigenfunctions of the quarks and
antiquarks.  These collision of produced particle will convert
longitudinal momenta to transverse momenta, leading to medium
particles with smaller longitudinal momenta but greater transverse
momenta in the approach to equilibrium.  Thirdly, the produced
particles will likely be subject to strong residual interactions after
production. The interaction is stronger, the more the strings are
closer packed together at higher collision energies.  Fourthly,
because of their mutual interaction and the nature of color
confinement, those produced particles reaching the boundary region will
form a surface in the boundary region, against which produced particle
arriving later will collide.  Part of these particles will continue to
propagate forward as a result of the collision with the longitudinal
boundary and will stretch the surface to a greater longitudinal
extension. However, part of the colliding produced particles will be
reflected from the surface and will stream in the backward direction,
replenishing some of the partons that have streamed backward, similar
to the description in Fig. 1 of \cite{Won82}, in the approach to
equilibrium.

We consider next the heavy-ion collisions at very high energies.  For
a collision energy of 100 GeV per nucleon at RHIC, the average
longitudinal separation between strings is of order 0.019 fm which is
much smaller than the width of the Wigner function, $\hbar/m_{\pi T}$.
The nucleus-nucleus collision leads to a large overlap of the
fragmenting strings.  Because of the strong overlap of the strings,
the produced quanta will propagate to the longitudinal boundary
together in nearly the same speed and nearly in unison.  Because of
the strong overlap and the Lorentz contraction, there is a large
density and large interaction among produced particles after they are
produced.  The produced particles will propagate according to their
longitudinal rapidities. Partons with large rapidities will reach the
boundary first, forming a surface region. The interacting partons near
the surface will collide with the confining and moving surface.  As a
result of the collision with the boundary, part of these partons will
continue forward, and they will stretch the surface boundary to a
greater longitudinal extension to keep it a continuously moving
boundary.  Because of the confining interactions at the moving
boundary, some of the other partons will come to the classical turning
point at the moving boundary.  They will then be reflected from the
surface and will flow in the backward direction.  The reflection of
confining partons leads to two important effects.  First, the
reflected partons will replenish some of the partons that have
streamed backward, as depicted in Fig. 1 of Ref.\ \cite{Won82}.
Secondly, the reflected partons will encounter partons streaming in
the opposite direction.  Such encountering will lead to collisions
which will convert longitudinal momentum to transverse momentum,
leading to medium particles with a smaller average longitudinal
momentum.  The collisions may relax the rapidity distribution from an
initial rapidity plateau \cite{Yan08} to a Gaussian rapidity
distribution at the end of nucleus-nucleus collision
\cite{Mur04,Ste05,Ste07}.  Whether or not such a scenario indeed
occurs will require more future analyses.

\section{Discussion and Conclusions}

We investigate the space-time behavior of produced particles in the
fragmentation of a string in QED2.  We obtain a relation between the
rapidity distribution and the Fourier transforms of the initial
fermion current.  We find that the rapidity plateau at high energies
arises as a consequence of a localized initial fermion current.  The
Wigner function of the produced particles show that particles with
momenta in different regions of the rapidity plateau are present at
the moment of string fragmentation, with the width of their spatial
distributions decreasing with an increasing center-of-mass energy.
The Wigner function exhibits the effects of interference in the
fragmentation of many identical strings.

Because QED2 mimics many features of QCD4, it is useful to examine the
circumstances in which a QCD4 systems can be approximated by QED2.  We
show first how QCD4 with transverse confinement can be approximately
compactified as QCD2 with a transverse quark mass $m_{{}_T}$ obtained
by solving a set of coupled transverse eigenvalue equations.
Furthermore, in the limit of the strong coupling and the large number
of flavors $N_f$, QCD2 admits Schwinger QED2-type solutions.  It is
for these reasons that QED2 can mimic many important features of QCD4,
including the properties of the proper high-energy rapidity plateau
behavior, quark confinement, charge screening, and chiral symmetry
breaking.  In the absence of rigorous non-perturbative QCD4 solutions,
it is therefore reasonable to study the particle production process
phenomenologically using Schwinger's QED2 model, with the boson quanta
of QED2 considered as analogous to the boson quanta in QCD.

There are however important differences which must be kept in mind in
the discussion of the dynamics subsequent to the fragmentation of
strings.  Free bosons are the quanta of QED2 while strongly
interacting gluons and quarks and relatively weakly interacting
hadrons are quanta of QCD4 at different temperatures of the QCD
system.  In a nucleus-nucleus collision at high energies, the
particles produced after the fragmentation the strings will be subject
to strong interactions with the dense medium of produced particles in
their vicinity after production.  These interactions are important as
the strings overlap strongly owing to the Lorentz contraction.  The
large number of colors and flavors make it likely that the overlapping
strings are of different types and do not interfere.  As a
consequence, the density of the produced particles accumulatively
increases as the density of strings increases.  It will be necessary
to include these strong interactions between the produced particles in
order to describe better the subsequent dynamics.  It will also be
necessary to restore the transverse degrees of freedom for a better
description by using information on the transverse eigenfunctions of
the quarks and antiquarks and their transverse momentum distributions.
Our formulation of the approximate compactification between
transversely-confined QCD4 into QCD2 facilitates such a restoration.

The main feature inferred from our present analysis is that in a
string fragmentation, boson field quanta with momenta in different
regions of the rapidity plateau are present in the initial Wigner
function at the moment of string fragmentation.  This peculiar
behavior arises from the quantum effects of the vacuum structure.
Particles in the vacuum of the Dirac sea participate in the
interaction as the valence fermions, antifermions, and gauge fields
change with space and time. The non-perturbative self-consistent
response of particles in the Dirac sea leads to the production of
boson quanta with momenta in different regions of the rapidity plateau
at the moment of string fragmentation, in contrast to the classical
picture of string fragmentation where there is a momentum-space-time
ordering of produced particles.

Our analysis of the dynamics of the Wigner function places it in the
class of initial-state-interaction description of quantum
chromodynamics.  Our result that partons in different regions of
rapidities over the rapidity plateau are produced at the moment of
collision is in line with many other well-justified
initial-state-interaction models, such as the parton model
\cite{Fey69}, the Drell-Yan process \cite{Dre71}, the multi-peripheral
model \cite{Che68}, and the dipole approach of photon-induced strong
interactions \cite{Huf00}.  These are models in which the constituent
particles or produced particles are present at or before the collision
process at $t=0$.
 
With regard to the momentum kick model which motivated the present
analysis, the above results may resolve one of the puzzles concerning
the possible occurrence of the particles with large longitudinal
momenta in the early stage of the nucleus-nucleus collisions.  Partons
of different rapidities are present in the initial parton momentum
distribution at fragmentation.  The jet produced in one of the
nucleon-nucleon collisions can find partons of large rapidities in the
early environment.  There can be collisions between the jet and these
high-rapidity partons, which show up as ridge particles in coincidence
with the jet.  Just as the presence of antiquark partons from the
quark-antiquark sea is revealed by the occurrence of the Drell-Yan
process in the initial-state-interaction picture, so it is here that
the presence of the rapidity plateau in the early parton momentum
distribution is revealed by the occurrence of large rapidity
associated particles in coincidence with the near-side jet in the
PHOBOS experiments.

\null
\vskip 0.5cm
\centerline{\bf Acknowledgment}
\vskip .5cm
The authors would like to thank Profs. H. W. Crater and Che-Ming Ko
for helpful discussions.  This research was supported in part by the
Division of Nuclear Physics, U.S. Department of Energy, under Contract
No.  DE-AC05-00OR22725, managed by UT-Battelle, LLC.


\begin{thebibliography}{99}
\bibitem{Aihnote} Please note that in Ref.\ \cite{Aih88}, the
experimental data of the rapidity distribution are presented as
$dN_{\rm ch}/d|y|$ which includes both positive and negative $y$
contributions, and $N_{\rm ch}$ is $N_{\pi^+ + \pi^-}$.  On the other
hand, the experimental data of the rapidity distribution are given as
$dN_{\rm ch}/dy$ for positive $y$ in Ref.\ \cite{Hof88}. They are
related by a factor of two as $dN_{\rm ch}/d|y|= 2 dN_{\rm ch}/dy$.
\bibitem{Not2} 
The signs on the righthand sides of in Eqs.\ (40) and (41) in Ref.\
\cite{Won08a} should be reversed as in Eqs. (\ref{dpdr1}) and
(\ref{dpdr2}) here.
\bibitem{Not1} The boson mass in the large $N_f$ limit of massless
multi-flavor QCD2 is given as $M^2=e^2N_f/2 \pi$ in
Ref.\  \cite{Arm99}, but as $M^2=e^2N_f/ \pi$ in
Refs. \cite{Tri02,Abr04}, using apprently different definitions of the
coupling constant $e$.  We shall follow the convention of
\cite{Tri02,Abr04} with the boson mass of $M^2=e^2N_f/ \pi$.
\end{thebibliography}
\end{document}